\documentclass[aps,preprint]{revtex4}%
\usepackage{amsfonts}
\usepackage{amsmath}
\usepackage{amssymb}
\usepackage{graphicx}%
\providecommand{\U}[1]{\protect\rule{.1in}{.1in}}
\newtheorem{theorem}{Theorem}
\newtheorem{acknowledgement}[theorem]{Acknowledgement}

\begin{document}
\preprint{UATP/1006}
\title{Non-equilibrium thermodynamics.\ IV: Generalization of Maxwell,
Claussius-Clapeyron and Response Functions Relations, and the Prigogine-Defay
Ratio for Systems\ in Internal Equilibrium}
\email{pdg@uakron.edu }
\author{P.D. Gujrati and P.P. Aung }
\affiliation{Department of Physics, Department of Polymer Science, The University of Akron,
Akron, OH 44325 USA}

\begin{abstract}
We follow the consequences of internal equilibrium in non-equilibrium systems
that has been introduced recently [Phys. Rev. E \textbf{81}, 051130 (2010)] to
obtain the generalization of Maxwell's relation and the Clausius-Clapeyron
relation that are normally given for equilibrium systems. The use of Jacobians
allow for a more compact way to address the generalized Maxwell relations; the
latter are available for any number of internal variables. The
Clausius-Clapeyron relation in the subspace of observables show not only the
non-equilibrium modification but also the modification due to internal
variables that play a dominant role in glasses. Real systems do not directly
turn into glasses (GL) that are frozen structures from the supercooled liquid
state L; there is an intermediate state (gL) where the internal variables are
not frozen. Thus, there is no single glass transition. A system possess
several kinds of glass transitions, some conventional (L$\rightarrow$gL;
gL$\rightarrow$GL) in which the state change continuously and the transition
mimics a \emph{continuous} or second order transition, and some apparent
(L$\rightarrow$gL; L$\rightarrow$GL) in which the free energies are
discontinuous so that the transition appears as a \emph{zeroth order}
transition, as discussed in the text. We evaluate the Prigogine-Defay ratio
$\Pi$ in the subspace of the observables at these transitions. We find that it
is normally different from $1,$ except at the conventional transition
L$\rightarrow$gL, where $\Pi=1$ regardless of the number of internal variables.

\end{abstract}
\date[December 15, 2010]{}
\maketitle

\section{Introduction}

\subsection{Previous Results}

\begin{figure}
[ptb]
\begin{center}
\includegraphics[
height=2.5322in,
width=5.2243in
]%
{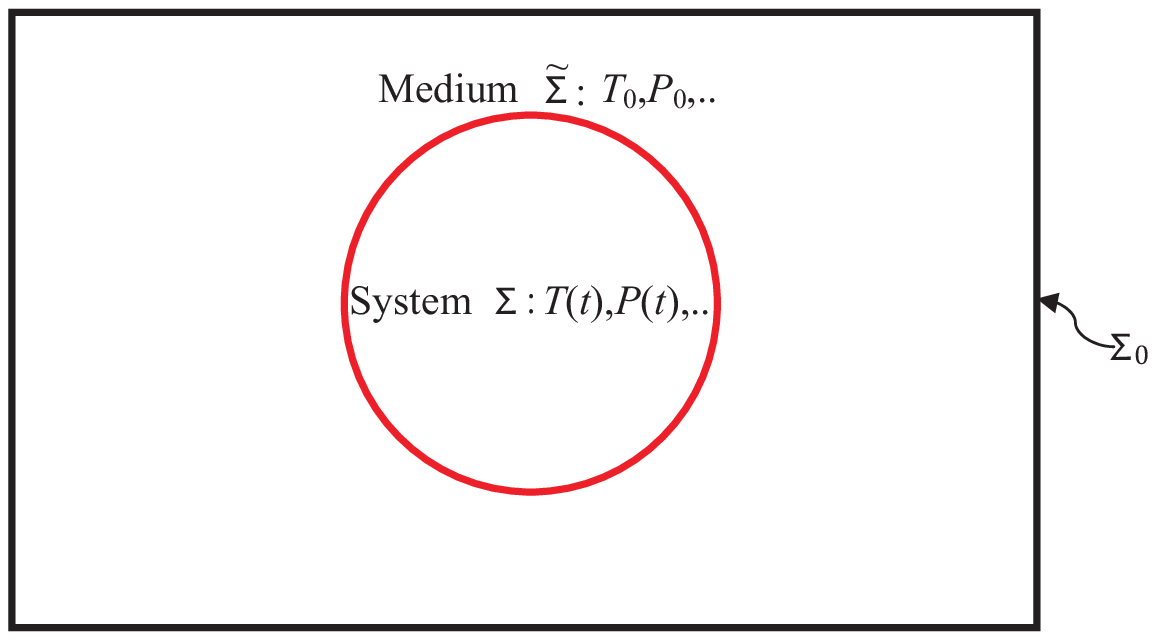}%
\caption{Schematic representation of a system $\Sigma$ and the medium
$\widetilde{\Sigma}$ surrounding it to form an isolated system $\Sigma_{0}$.
The medium is described by its fields $T_{0},P_{0},$ etc. while the system, if
in internal equilibrium (see text) is characterized by $T(t),P(t),$ etc.}%
\label{Fig_Systems}%
\end{center}
\end{figure}

In a series of papers, we have begun to develop non-equilibrium thermodynamics
starting from the second law and ensuring the additivity of entropy as a state
function
\cite{Gujrati-Non-Equilibrium-I,Gujrati-Non-Equilibrium-II,Gujrati-Non-Equilibrium-III}%
. The central idea in this approach is that of \emph{internal equilibrium}
within a macroscopic system $\Sigma$ surrounded by an extremely large medium
$\widetilde{\Sigma}$; the two form an isolated system $\Sigma_{0}$ as shown in
Fig. \ref{Fig_Systems}. While the entropy $S(t)$ and the general
non-equilibrium thermodynamic potential $\Omega(t)$, see
\cite{Gujrati-Non-Equilibrium-II} for more details, such as the
non-equilibrium Gibbs free energy $G(t)$ of the system \emph{exist} even when
the system is not in internal equilibrium, the Gibbs fundamental relation
exists only when the system is in internal equilibrium:%
\begin{equation}
dS(t)=\mathbf{y}(t)\mathbf{\cdot}d\mathbf{X}(t)\mathbf{+a(t)\cdot}%
d\mathbf{I}(t)\mathbf{,} \label{Gibbs_Fundamental_relation_0}%
\end{equation}
where $\mathbf{X}(t)\mathbf{\ }$and $\mathbf{I}(t)$ represent the set of
observables and the set of internal variables, respectively, to be
collectively denote by $\mathbf{Z}(t)$. The entropy $S(\mathbf{Z}(t),t)$ away
from equilibrium, no matter how far from equilibrium, is normally a function
of $\mathbf{Z}(t)$ and $t$. However, when the system is in internal
equilibrium, where Eq. (\ref{Gibbs_Fundamental_relation_0}) remains valid,
$S(t)$ has \emph{no} explicit $t$-depenedence; the temporal evolution of the
entropy in this case comes from the time-dependence in $\mathbf{Z}(t)$, with
$\mathbf{X}(t)\mathbf{\ }$and $\mathbf{I}(t)$ still independent of each other.
The coefficient $\mathbf{y}(t)\mathbf{\ }$and $\mathbf{a}(t)$ represent the
derivatives of the entropy and are normally called the internal field and the
internal affinity, respectively. The energy $E$, volume $V$ and the number of
particles $N$ play a very special role among the observables, and the
corresponding internal fields are given by%
%TCIMACRO{\FRAME{ftbpFU}{5.2243in}{2.5322in}{0pt}{\Qcb{Schematic representation
%of a system $\Sigma$ and the medium $\widetilde{\Sigma}$ surrounding it to
%form an isolated system $\Sigma_{0}$. The medium is described by its fields
%$T_{0},P_{0},$ etc. while the system, if in internal equilibrium (see text) is
%characterized by $T(t),P(t),$ etc.}}{\Qlb{Fig_Systems}}{system_modified_1.eps}%
%width 5.2243in;  height 2.5322in;  depth 0pt;  original-width 5.1673in;
%original-height 2.4915in;  cropleft "0";  croptop "1";  cropright "1";
%cropbottom "0";  filename 'System_Modified_1.eps';file-properties "XNPEU";}}}%
%BeginExpansion
%EndExpansion
%

\begin{equation}
\frac{1}{T(t)}=\left(  \frac{\partial S(t)}{\partial E(t)}\right)
_{\mathbf{Z}^{\prime}(t)},\ \ \frac{P(t)}{T(t)}=\left(  \frac{\partial
S(t)}{\partial V(t)}\right)  _{\mathbf{Z}^{\prime}(t)},\ \frac{\mu(t)}%
{T(t)}=-\left(  \frac{\partial S(t)}{\partial N(t)}\right)  ,
\label{Field_Variables_0}%
\end{equation}
where $\mathbf{Z}^{\prime}(t)$ denotes all other elements of $\mathbf{Z}(t)$
except the one used in the derivative. Thus, internal temperature, pressure,
etc. have a meaning only when the system comes into internal equilibrium. In
general, the internal field $\mathbf{y}(t)$ and affinity $\mathbf{a}(t)$ are
given by%
\begin{equation}
\mathbf{y}(t)\equiv\frac{\mathbf{Y}(t)}{T(t)}\equiv\left(  \frac{\partial
S(t)}{\partial\mathbf{X}(t)}\right)  _{\mathbf{Z}^{\prime}(t)},\mathbf{a}%
(t)\equiv\frac{\mathbf{A}(t)}{T(t)}\equiv\left(  \frac{\partial S(t)}%
{\partial\mathbf{I}(t)}\right)  _{\mathbf{Z}^{\prime}(t)}.
\label{Field_Variables_1}%
\end{equation}
The fields of the medium $T_{0},P_{0,}\mu_{0}$, etc., which we collectively
denote by $\mathbf{Y}_{0}$, are different from the internal fields of the
system unless the latter comes to equilibrium with the medium. The same is
also true of the affinity, except that the affinity vector $\mathbf{A}_{0}=0$
for the medium; see II.

From now on, we will only consider the case when the system is in internal
equilibrium. The heat transfer is given by
\begin{equation}
dQ=T(t)dS(t)=T_{0}d_{\text{e}}S(t), \label{Heat_Transfer}%
\end{equation}
where $d_{\text{e}}S(t)$ is the entropy exchange with the medium. The
irreversible entropy generation $d_{\text{i}}S(t)$ within the system is given
by%
\[
d_{\text{i}}S(t)\equiv dS(t)-d_{\text{e}}S(t)\geq0.
\]
Thus, as long as the system is not in equilibrium, $T(t)\neq T_{0};$
accordingly, $d_{\text{i}}S(t)>0$ in accordance with the second law. There is
irreversible entropy production even when the system is in internal
equilibrium; the latter only allows us to introduce the internal fields and
affinities via Eqs. (\ref{Field_Variables_0}) and (\ref{Field_Variables_1}).

In the absence of any internal variables, the Gibbs fundamental relation is
given by
\begin{equation}
T(t)dS(t)=dE(t)+P(t)dV(t)-\mu(t)dN(t)
\label{Gibbs_Fundamental_relation_Internal}%
\end{equation}
for the special case when $\mathbf{X}(t)$ only contains $E(t),V(t)$ and
$N(t)$. For a fixed number of particles, the last term would be absent. As
said above, the temperature, pressure, etc. of the medium and the system are
usually different when the system is out of equilibrium with the medium. Only
in equilibrium do they become equal, in which case, the Gibbs fundamental
relation in Eq. (\ref{Gibbs_Fundamental_relation_Internal}) reduces to the
standard form%
\begin{equation}
T_{0}dS_{\text{eq}}=dE_{\text{eq}}+P_{0}dV_{\text{eq}}-\mu_{0}dN_{\text{eq}},
\label{Gibbs_Fundamental_relation_Equilibrium}%
\end{equation}
in which none of the quantities has any time-dependence; the extensive
quantities represent the equilibrium values and are denoted by the additional
suffix. One normally considers a system with fixed number of particles, in
which case, the last term is absent in Eq.
(\ref{Gibbs_Fundamental_relation_Equilibrium}). In the following, we will not
explicitly show the additional suffix unless clarity is needed. The following
Maxwell relations that follow from the Gibbs fundamental relation, see Eq.
(\ref{Gibbs_Fundamental_relation_Equilibrium}), are well-known and can be
found in any good text-book such as \cite{Landau}\ on thermodynamics:%

\begin{align}
\left(  \frac{\partial T_{0}}{\partial V}\right)  _{S,N}  &  =-\left(
\frac{\partial P_{0}}{\partial S}\right)  _{V,N},\ \ \ \left(  \frac{\partial
T_{0}}{\partial P_{0}}\right)  _{S,N}=\left(  \frac{\partial V}{\partial
S}\right)  _{P_{0},N},\nonumber\\
\left(  \frac{\partial P_{0}}{\partial T_{0}}\right)  _{V,N}  &  =\left(
\frac{\partial S}{\partial V}\right)  _{T_{0},N},\ \ \ \left(  \frac{\partial
S}{\partial P_{0}}\right)  _{T_{0},N}=-\left(  \frac{\partial V}{\partial
T_{0}}\right)  _{P_{0},N}. \label{Maxwell_Relations}%
\end{align}
In equilibrium, there is no explicit $t$-dependence in $\mathbf{Z}$; moreover,
the internal variable $\mathbf{I}$ is no longer independent of $\mathbf{X}$.
The equilibrium field and affinity of the system become equal to those of the
medium ($\mathbf{Y}_{0}$ and $\mathbf{A}_{0}=0$); see
\cite{Gujrati-Non-Equilibrium-II}. Thus, the Gibbs fundamental relation
reduces to
\begin{equation}
dS=\mathbf{y}_{0}\mathbf{\cdot}d\mathbf{X},
\label{Gibbs_Fundamental_relation_Equilibrium_0}%
\end{equation}
compare with Eq. (\ref{Gibbs_Fundamental_relation_Equilibrium}). The
equilibrium value of the internal variable can be expressed as a function of
the equilibrium value of $\mathbf{X}$:%
\[
\mathbf{I=I}_{\text{eq}}(\mathbf{X}_{\text{eq}}).
\]

We now observe the similarity between the Gibbs fundamental relations in Eqs.
(\ref{Gibbs_Fundamental_relation_Equilibrium}) and
(\ref{Gibbs_Fundamental_relation_Equilibrium_0}). This strongly suggests that
there may also exist analogs of the Maxwell relations or other important
relations that are based on Eq.
(\ref{Gibbs_Fundamental_relation_Equilibrium_0}) for a system that, although
not in equilibrium with the medium, is in internal equilibrium. In this sequel
to the earlier papers
\cite{Gujrati-Non-Equilibrium-I,Gujrati-Non-Equilibrium-II,Gujrati-Non-Equilibrium-III}%
, which we denote by I, II, and III, respectively, we develop the consequence
of this internal equilibrium thermodynamics for important relations such as
Maxwell relations, Clausius-Clayperon equation, etc. These extensions will
play an important role in non-equilibrium systems that are nonetheless in
internal equilibrium.

The time-variation of the internal temperature $T$ of a non-equilibrium system
such as a glass is due to the time dependence of the observable $\mathbf{X}%
(t)$ such as $E(t),V(t)$, etc. and of the internal variable $\mathbf{I}(t)$.
For example, at fixed $T_{0}$, the internal temperature will continue to
change during structural relaxation. The internal temperature will also change
if the temperature of the medium changes. Thus%
\[
dT=\left(  \frac{\partial T}{\partial\mathbf{X}}\right)  \cdot d\mathbf{X}%
+\left(  \frac{\partial T}{\partial\mathbf{I}}\right)  \cdot d\mathbf{I.}%
\]
The rate of change of the internal temperature can be expressed in terms of
the rate of change $r=dT_{0}/dt:$%
\begin{equation}
\frac{dT}{dt}=\left(  \frac{\partial T}{\partial\mathbf{X}}\right)  \cdot
\frac{d\mathbf{X}}{dt}+\left(  \frac{\partial T}{\partial\mathbf{I}}\right)
\cdot\frac{d\mathbf{I}}{dt}\mathbf{.} \label{temperature-time_derivative}%
\end{equation}
Similarly,%
\begin{equation}
\frac{dT}{dT_{0}}=\left(  \frac{\partial T}{\partial\mathbf{X}}\right)
\cdot\frac{d\mathbf{X}}{dT_{0}}+\left(  \frac{\partial T}{\partial\mathbf{I}%
}\right)  \cdot\frac{d\mathbf{I}}{dT_{0}}. \label{T_T0_Derivative0}%
\end{equation}
The same analysis can be carried out for other internal fields.

\subsection{Present Goal}

Our aim in this work is to follow the consequences of internal equilibrium in
a non-equilibrium system to find the generalization of Maxwell's relations,
the Clausius-Clapeyron relation, and the relations between response functions
to non-equilibrium states. We will be also be interested in glasses in this
work; they are traditionally treated as non-equilibrium states. Therefore, we
begin with a discussion of what is customarily called a glass and the
associated glass transition in Sect. \ref{Marker_Glass_Transitions}. A careful
discussion shows that the term does not refer to one single transition;
rather, it can refer to different kinds of transitions, some of which appear
similar to the conventional transitions in equilibrium, but the other refer to
apparent transitions where the Gibbs free energy cannot be continuous. There
are some well-known approximate approaches to glasses. We will briefly discuss
them. We then turn to our main goal to extend the Maxwell's relations, where
Jacobians are found to be quite useful. Therefore, we introduce Jacobians and
their various important properties in Sect. \ref{Marker_Jacobians}. This is
technical section, but we provide most of the required details so that the
clarity of presentation is not compromised. An important part of this section
is to show that the Jacobians can be manipulated in a straight forward manner
even in a subspace of the variables. This is important as the observations
require manipulating the observables and not the internal variables. Thus, the
experimental space refers to a subspace (Sect. \ref{Marker_Subspace}) of the
space where non-equilibrium thermodynamics is developed. Thermodynamic
potentials for non-equilibrium states are formulated in Sect.
\ref{Marker_Potentials}. We develop the generalization of the Maxwell's
relations in Sect. \ref{Marker_Maxwell_Relations}. We discuss generalization
of the Clausius-Clapeyron relation in Sect.
\ref{Marker_Clausius_Clapeyron_Relation}, where we also discuss the conditions
for phase transitions in non-equilibrium states. The response functions such
as the heat capacities, compressibilities and the expansion coefficients and
various relations among them are developed for non-equilibrium states in Sect.
Sect. \ref{Marker_Response_Functions}. The Prigogine-Defay ratio for glasses
are evaluated at various possible glass transitions in Sect.
\ref{Marker_PD_Ratio}. We compare our approach with some of the existing
approaches in determining the ratio in this section. The last section contains
a brief summary of our results.

\section{Glass Transitions and Apparent Glass
Transitions\label{Marker_Glass_Transitions}}

\begin{figure}
[ptb]
\begin{center}
\includegraphics[
trim=0.595673in 0.000000in -0.099027in 0.700200in,
height=3.2093in,
width=4.9338in
]%
{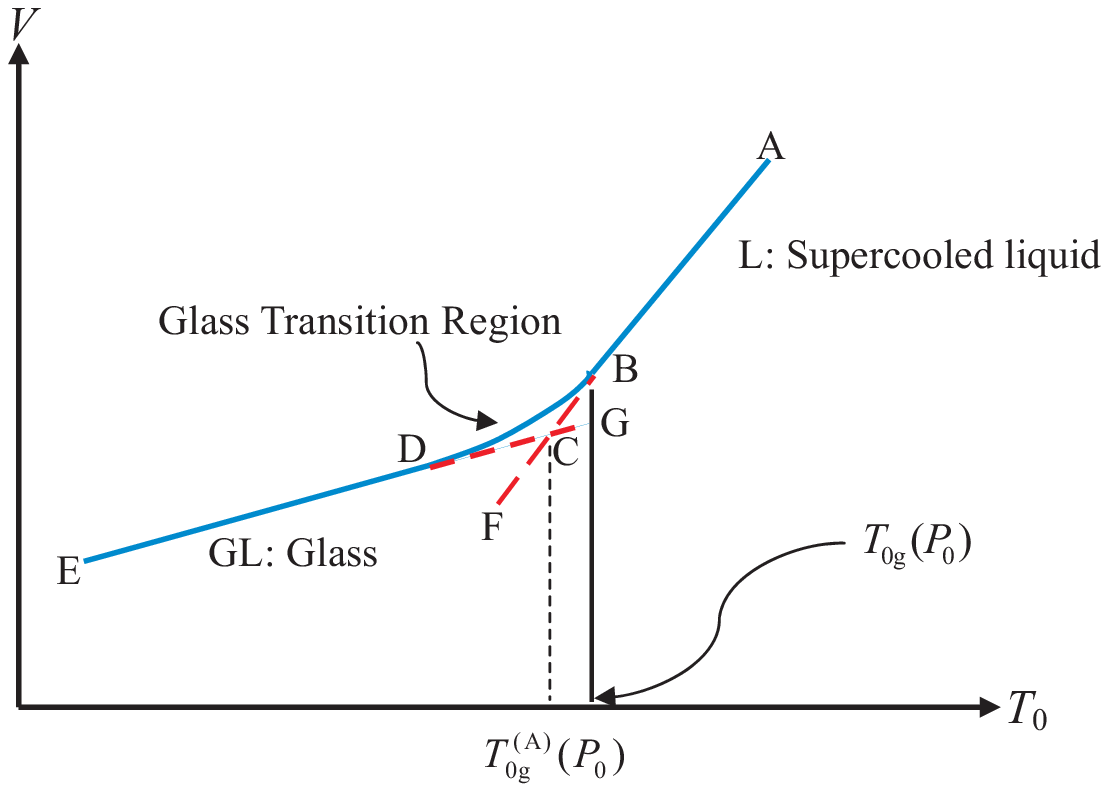}%
\caption{Schematic form of isobaric $V$ as a function of $T_{0}$ for a given
cooling rate. The pressure is fixed at $P_{0}$. The supercooled liquid turns
gradually into a glass through the glass transition region. The transition
temperature $T_{0\text{g}}(P_{0})$ is identified as the temperature at B,
where the actual volume begins to deviate from the extrapolated supercooled
liquid volume BC. On the other hand, the apparent glass transition temperature
$T_{0\text{g}}^{(\text{A})}(P_{0})$ is the temperature where the extrapolated
glass volume DC\ meets the extrapolated supercooled liquid volume BC as
indicated in the fuigure; this temperature lies in the glass transition
region. }%
\label{Fig_GlassTransition_V}%
\end{center}
\end{figure}

\begin{figure}
[ptb]
\begin{center}
\includegraphics[
trim=0.499262in 0.000000in 0.000000in 0.599710in,
height=3.4705in,
width=5.041in
]%
{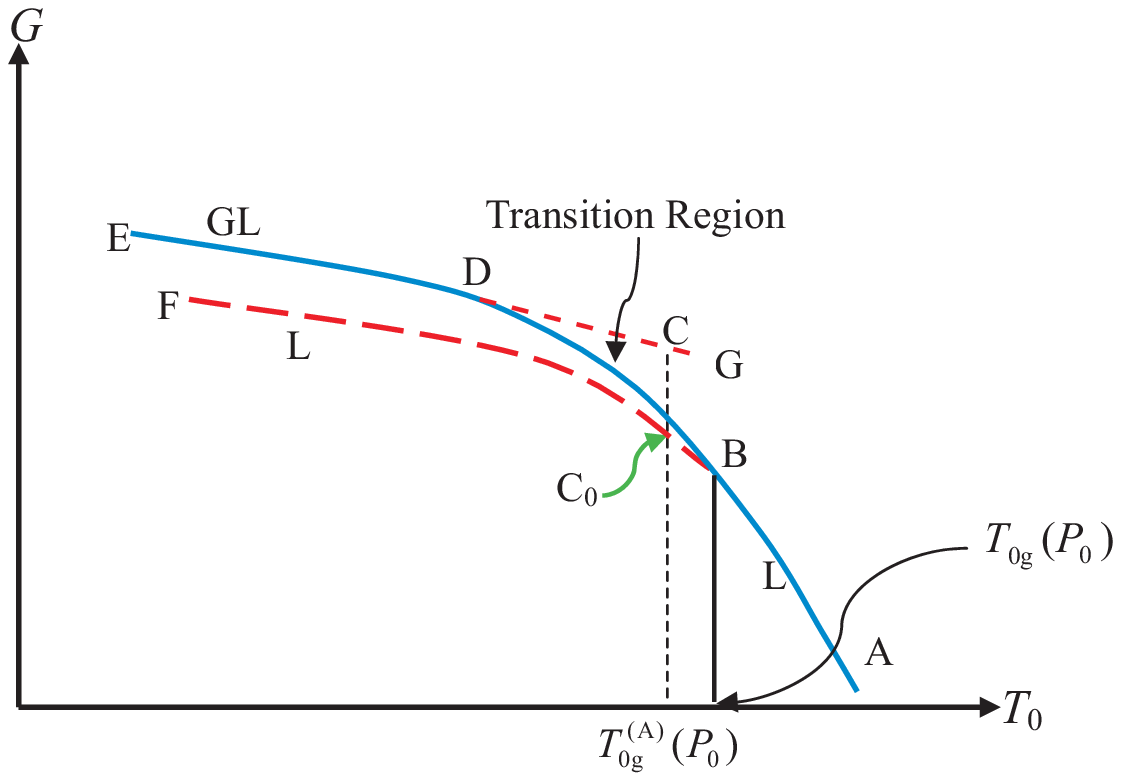}%
\caption{Schematic form of the isobaric Gibbs free energy $G$ shown by the
continuous curve ABDE as a function of the medium temperature $T_{0}$ at a
fixed pressure $P_{0}$. The extrapolation of the glassy portion (GL) along DCG
and the supercooled liquid (L) portion ABC$_{0}$F do not meet; the glassy
Gibbs free energy at point the apparent glass transition C, where
$T_{0}=T_{0\text{g}}^{(\text{A})}(P_{0})$, is higher than that at
C$_{\text{0}}$ on the continuous curve L at the same temperature
$T_{0\text{g}}^{(\text{A})}(P_{0})$, showing that the extrapolation results in
a more unstable state at the apparent glass transition C than the physical
state C$_{\text{0}}$ on the continuous curve. The Gibbs free energies match at
the glass transition temperature $T_{0\text{g}}(P_{0})$ at B$.$ }%
\label{Fig_GlassTransition_G}%
\end{center}
\end{figure}

An example of non-equilibrium systems under investigation here is a glass
\cite{Gujrati-book}; see Figs. \ref{Fig_GlassTransition_V} and
\ref{Fig_GlassTransition_G}. A supercooled liquid L is a \emph{stationary}
(time-independent) metastable state \cite{Gujrati-book}, which for our
purpose, represents an equilibrium state (by not allowing the crystalline
state into consideration), and is shown by the curve ABF under isobaric
condition at a fixed pressure $P_{0}$ of the medium. We will refer to the
equilibrium liquid always as L in the following. In contrast, a
non-equilibrium liquid state will be designated gL here, and represents a
\emph{time-dependent} metastable state \cite{Gujrati-book}. The choice of gL
is to remind us that it is a precursor to the eventual glass GL at a lower
temperature. The equilibrium liquid L is obtained by cooling the liquid L and
waiting long enough at each for it to come to equilibrium with the medium.
However, if it is obtained at a fixed cooling rate $r$, then at some
temperature $T_{0\text{g}}(P_{0}),$ L cannot come to equilibrium and turns
into gL; the resulting curve BD leaves ABF tangentially at B, and gradually
turns into an isobaric glass GL represented by the segment DE at D, when the
viscosity becomes so large ($\sim10^{13}$ poises) that it appears as a solid.
At B, the transition is from an equilibrium liquid L to a non-equilibrium
liquid form gL, and will be called the L-gL transition. In the literature, it
is commonly known as a transition from an ergodic state (L) to a non-ergodic
state (gL). In our opinion, this is a misnomer, as the concept of
\emph{ergodicity} refers to the long-time , indeed the infinite-time,
behavior. In this limit, there will be no gL, only L. Therefore, we will refer
to this transition at $T=T_{0\text{g}}(P_{0})$ as a L-gL transition or a
\emph{precursory} glass transition. The true glass transition at D is not a
transition from L to GL, but a transition from gL to GL. We will refer to the
glass transition at the lower temperature $T_{0\text{G}}(P_{0})$ at D as the
\emph{actual} glass transition, or simply the glass transition. The transition
region BD represents a time-dependent metastable supercooled liquid (to be
distinguished from the stationary metastable supercooled liquid L denoted by
ABF), which turns into a glass at D. The expansion coefficient in the glass is
almost identical to that of the corresponding crystal below D. The glass
continuously emerges out of gL at D, whose location is also determined by the
rate $r$ of cooling. The relaxation time $\tau$ of the system (the supercooled
liquid) becomes equal (really comparable) to the observation time
$\tau_{\text{obs}}$ at B. As seen in Fig. \ref{Fig_GlassTransition_V}, the
volume remains continuous at B and D at the two glass transitions. The same is
also true of the entropy. Indeed, the state of the system changes continuously
at B and D, which is highly reproducible for a given cooling rate $r$ or the
observation time $\tau_{\text{obs}}$. Thus, the points B and D can be taken as
a well-defined and unique glass transition temperatures $T_{0\text{g}}(P_{0})$
and $T_{0\text{G}}(P_{0})$ associated with the point B and D, respectively, in
both figures. Both transitions represent a non-equilibrium version of a
continuous transition (See Sect. \ref{Marker_Clausius_Clapeyron_Relation} for
elaboration on this point), where not only the Gibbs free energy, see Fig.
\ref{Fig_GlassTransition_G}, but also its derivatives are continuous. The
non-equilibrium nature of the transition appears in the dependence of the
value of $T_{0\text{g}}(P_{0})$ and $T_{0\text{G}}(P_{0})$ on the rate of
cooling. The continuity of the Gibbs free energy at B and D makes them as
genuine candidates as (glass) transition points, a requirement of a transition
in equilibrium thermodynamics. Therefore, both these transitions will be
collectively called \emph{conventional transitions} in this work.%
%TCIMACRO{\FRAME{ftbpFU}{4.9338in}{3.2093in}{0pt}{\Qcb{Schematic form of
%isobaric $V$ as a function of $T_{0}$ for a given cooling rate. The pressure
%is fixed at $P_{0}$. The supercooled liquid turns gradually into a glass
%through the glass transition region. The transition temperature $T_{0\text{g}%
%}(P_{0})$ is identified as the temperature at B, where the actual volume
%begins to deviate from the extrapolated supercooled liquid volume BC. On the
%other hand, the apparent glass transition temperature $T_{0\text{g}%
%}^{(\text{A})}(P_{0})$ is the temperature where the extrapolated glass volume
%DC\ meets the extrapolated supercooled liquid volume BC as indicated in the
%fuigure; this temperature lies in the glass transition region. }%
%}{\Qlb{Fig_GlassTransition_V}}{glasstransition_v.eps}%
%width 4.9338in;  height 3.2093in;  depth 0pt;  original-width 7.5593in;
%original-height 4.0733in;  cropleft "0.0788";  croptop "0.8281";
%cropright "1.0131";  cropbottom "0";
%filename 'GlassTransition_V.eps';file-properties "XNPEU";}}}%
%BeginExpansion
%EndExpansion

Unfortunately, the idea of a glass transition was formulated as a transition
between L and GL. Thus, neither of the above two glass transitions represent
the glass transition in the original sense. As the glass is considered a
frozen state, it is common to assume that over the region DE, the glass has
its internal variables denoted by $\mathbf{I}$ frozen at its value
$\mathbf{I}_{\text{G}}$ at D, even though its observables denoted by
$\mathbf{X}$ continue to change. On the other hand, the internal variables and
the observables continue to change over BD from their values at B to their
values at D. Consequently,
%TCIMACRO{\FRAME{ftbpFU}{5.041in}{3.4705in}{0pt}{\Qcb{Schematic form of the
%isobaric Gibbs free energy $G$ shown by the continuous curve ABDE as a
%function of the medium temperature $T_{0}$ at a fixed pressure $P_{0}$. The
%extrapolation of the glassy portion (GL) along DCG and the supercooled liquid
%(L) portion ABC$_{0}$F do not meet; the glassy Gibbs free energy at point the
%apparent glass transition C, where $T_{0}=T_{0\text{g}}^{(\text{A})}(P_{0})$,
%is higher than that at C$_{\text{0}}$ on the continuous curve L at the same
%temperature $T_{0\text{g}}^{(\text{A})}(P_{0})$, showing that the
%extrapolation results in a more unstable state at the apparent glass
%transition C than the physical state C$_{\text{0}}$ on the continuous curve.
%The Gibbs free energies match at the glass transition temperature
%$T_{0\text{g}}(P_{0})$ at B$.$ }}{\Qlb{Fig_GlassTransition_G}}%
%{glasstransition_gibbsfreeenergy.eps}{\special{ language "Scientific Word";
%type "GRAPHIC";  maintain-aspect-ratio TRUE;  display "USEDEF";
%valid_file "F";  width 5.041in;  height 3.4705in;  depth 0pt;
%original-width 5.4864in;  original-height 4.0222in;  cropleft "0.0910";
%croptop "0.8509";  cropright "1";  cropbottom "0";
%filename 'GlassTransition_GibbsFreeEnergy.eps';file-properties "XNPEU";}}}%
%BeginExpansion
%EndExpansion
the properties such as the volume of gL, which is shown schematically in Fig.
\ref{Fig_GlassTransition_V}, gradually change to those of the glass at lower
temperatures. Thus, the glass transition from AB\ to DE is not a sharp
transition. It can be argued, as we have done above, that B and D should be
taken as the glass transition points. However, the practice in the field is to
take a point between BD as a transition point obtained by electing some
well-defined rule of selection; see for example \cite{Gutzow} for a good
discussion of various ways of identifying the glass transition temperature.
One such rule commonly used is to consider the volume of the system and
introduce an \emph{apparent }glass transition temperature $T_{0\text{g}%
}^{(\text{A})}(P_{0})$ by the \emph{equilibrium continuation} of the volume
BCF of AB and by the \emph{extrapolation }of the volume DCG of DE to find
their crossing point C. The state of the glass following Tool and
Narayanaswamy \cite{Tool,Narayanaswamy} is then \emph{customarily} identified
by the point C on DC. However, there is no reason to take the state at C to
represent any real glass, as the extrapolation does not have to satisfy
non-equilibrium thermodynamics; the latter is valid only along the physical
path DB for the \emph{given history }of preparation such as determined by the
fixed rate $r$ of cooling during vitrification. The glass at $T_{0\text{g}%
}^{(\text{A})}(P_{0})$ must be described by the point on DB corresponding to
$T_{0\text{g}}^{(\text{A})}(P_{0})$ if we wish to employ non-equilibrium
thermodynamics. To be sure, one can find a slow enough cooling rate than the
one used to obtain gL at B so that the point B actually coincides with the
point C on ABF, as the latter represents L. However, the gL that will emerge
at C for the slower cooling rate has nothing to do with the extrapolated state
C on DCG. Because of the continuity of the state, the gL at the slower rate at
C will have its $A=0$ and $\xi=\xi_{\text{eq}}~$\ and will have its Gibbs free
energy continuous. Moreover, the new gL will follow a curve that will be
strictly below BDE. These aspects make the new gL different from the
extrapolated GL\ at C. Taking the point C on CD to represent the glass will be
an approximation, which we will avoid in this work, as our interest is to
apply thermodynamics in the study of glasses. Therefore, we will use the
extrapolation to only determine $T_{0\text{g}}^{(\text{A})}(P_{0})$, but the
real glass and the real liquid states are determined by the curve BD and BCF,
respectively, where our non-equilibrium thermodynamics should be applicable.

The location of this temperature $T_{0\text{g}}^{(\text{A})}(P_{0})$ depends
on the property being extrapolated. We can use the entropy of the system to
locate the apparent glass transition temperature, which would invariably give
a different value for the apparent glass transition temperature. To call one
of these temperatures as a transition temperature is a misnomer for another
reason. None of these temperatures represent a "non-equilibrium" thermodynamic
transition for the simple reason that the two branches DCG and BCF do not have
a common Gibbs free energy \ at $T_{0\text{g}}^{(\text{A})}(P_{0})$ as is
clearly seen in Fig. \ref{Fig_GlassTransition_G}. The branch ABC$_{0}$F
represents the Gibbs free energy of the equilibrium supercooled liquid, while
the segment DE represents the Gibbs free energy of the glass, with the segment
BD denoting the Gibbs free energy of the system during the transition region.
The extrapolation DCG in Fig. \ref{Fig_GlassTransition_V} to determine the
glass transition temperature $T_{0\text{g}}^{(\text{A})}(P_{0})$ corresponds
to the extrapolated segment DCG in Fig. \ref{Fig_GlassTransition_G}. The Gibbs
free energy of the glass in this extrapolation is given by the point C, while
the Gibbs free energy of the supercooled liquid is determined by the point
C$_{0}$. Evidently, the two free energies are very different, with that of the
glass higher than that of the supercooled liquid, as expected from the
non-equilibrium nature of the glassy state.

The above discussion of the apparent glass transition also applies to
comparing the glass at D with the corresponding L at $T_{0\text{G}}(P_{0})$,
which will represent yet another apparent glass transition temperature. This
apparent glass transition has the same problem regarding the Gibbs free energy
as the previous one at $T_{0\text{g}}^{(\text{A})}(P_{0})$. However, this
transition differs from the apparent glass transition at $T_{0\text{g}%
}^{(\text{A})}(P_{0})$\ in that the "glass" at $T_{0\text{g}}^{(\text{A}%
)}(P_{0})$\ is not a frozen state, while the glass at D is a "frozen" glass to
some extent (as it also undergoes structural relaxation in time). It should
also be remarked that whether we consider the apparent glass transition at
$T_{0\text{G}}(P_{0})$ or $T_{0\text{g}}^{(\text{A})}(P_{0})$, the transition
is an example of a discontinuity in the Gibbs free energy of the two states.
This is different from the precursory glass transition and the actual glass
transition at B and D, respectively, where the Gibbs free energy is
continuous. Because of the discontinuity in the Gibbs free energies in the
apparent glass transitions at $T_{0\text{G}}(P_{0})$ and $T_{0\text{g}%
}^{(\text{A})}(P_{0})$, we will refer to these transitions as \emph{apparent
transitions} in this work. Indeed, one can think of these transitions as an
analog of a\emph{ zeroth order} transition because of the discontinuity in the
Gibbs free energy. However, it should be remarked that the apparent
transitions do not represent any transition in the system; those transitions
are the two conventional transitions discussed above. The apparent transitions
represent our desire to compare two distinct states. This is like comparing
the supercooled liquid with the crystal at the same temperature and pressure.
Therefore, a discontinuity in the Gibbs free energy is not a violation of the
principle of continuity discussed in \cite{Gujrati-Non-Equilibrium-III}.

We will consider all of the above glass transitions later when we discuss the
evaluation of the Prigogine-Defay ratio
\cite{Prigogine-Defay,Davies,Goldstein,Gupta,DiMarzio} in Sect.
\ref{Marker_PD_Ratio}. In this ratio, a non-equilibrium state is compared with
the equilibrium supercooled liquid state along ABF. In the classic approach
adopted by Simon \cite{Simon,Gutzow}, the temperature range $(T_{0\text{gG}%
}(P_{0}),T_{0\text{g}}(P_{0}))$ is shrunk to a point, either by considering
the apparent glass transition at $T_{0\text{g}}^{(\text{A})}(P_{0})$, or by
comparing the glass state at D with the supercooled liquid L at B. The latter
amounts to neglecting the segment BD from consideration. We will avoid this
ad-hoc approach in this work. The only possible scenario, where Simon's
approach is meaningful is that of the \emph{ideal glass transition }\cite[and
references thererin]{Gujrati-book}, in the limit the cooling rate
$r\rightarrow0$. In this limiting case, the crossover region BD disappears and
the ideal glass IGL emerges directly out of the L at the ideal glass
transition temperature $T_{0\text{IG}}$. This is a conventional continuous
transition between the two stationary states IGL and L, both of which remain
in equilibrium with the medium at $T_{0},P_{0}$. There is no need to invoke
any internal variable $\mathbf{I}$ to describe the ideal glass; the observable
$\mathbf{X}$ is sufficient for the investigation of the ideal glass
transition. We will revisit this point later in Sect. \ref{Marker_PD_Ratio}.

\section{Some Useful Mathematical Tools}

\subsection{Jacobian method\label{Marker_Jacobians}}

Jacobians \cite{Courant} will be found extremely useful in this work just as
they are found useful in equilibrium thermodynamics \cite{Landau}; see also
\cite{Shaw,Crawford,Pinkerton}. The $n$-th order Jacobian of $u_{1}%
,u_{2},\cdots u_{n}$ with respect to $x_{1},x_{2},\cdots x_{n}$ is the
$n\times n$ determinant of the matrix formed by $\partial u_{k}/\partial
x_{l}$:%
\[
\frac{\partial(u_{1},u_{2},\cdots u_{n})}{\partial(x_{1},x_{2},\cdots x_{n}%
)}\equiv\left\vert
\begin{array}
[c]{ccccc}%
\partial u_{1}/\partial x_{1} & \partial u_{1}/\partial x_{2} & . & . &
\partial u_{1}/\partial x_{n}\\
\partial u_{2}/\partial x_{1} & \partial u_{2}/\partial x_{2} & . & . &
\partial u_{2}/\partial x_{n}\\
. & . & . & . & .\\
. & . & . & . & .\\
\partial u_{n}/\partial x_{1} & \partial u_{n}/\partial x_{2} & . & . &
\partial u_{n}/\partial x_{n}%
\end{array}
\right\vert .
\]
It is clear from the properties of the determinant that

\begin{enumerate}
\item The Jacobian vanishes if any two $u$'s are identical%
\[
\frac{\partial(u_{1},u_{2},\cdots u_{i},u_{i}\cdots u_{n})}{\partial
(x_{1},x_{2},\cdots x_{i},x_{i+1}\cdots x_{n})}=0.
\]

\item If $u_{i}$ and $u_{i+1}$ interchange their order, the Jacobian changes
its sign%
\[
\frac{\partial(u_{1},u_{2},\cdots u_{i+1},u_{i}\cdots u_{n})}{\partial
(x_{1},x_{2},\cdots x_{i},x_{i+1}\cdots x_{n})}=-\frac{\partial(u_{1}%
,u_{2},\cdots u_{i},u_{i+1}\cdots u_{n})}{\partial(x_{1},x_{2},\cdots
x_{i},x_{i+1}\cdots x_{n})}.
\]

\item If any $u_{i}$ is equal to $x_{i}$, the $n$-th order Jacobian reduces to
a $(n-1)$-th order Jacobian formed by derivatives at fixed $x_{i}$. For
example, for $n=2$, we have%
\[
\frac{\partial(u_{1},x_{2})}{\partial(x_{1},x_{2})}=\left(  \frac{\partial
u_{1}}{\partial x_{1}}\right)  _{x_{2}}.
\]

\end{enumerate}

When we consider compound transformations $\left(  x_{1},x_{2},\cdots
x_{n}\right)  \rightarrow\left(  u_{1},u_{2},\cdots u_{n}\right)
\rightarrow\left(  v_{1},v_{2},\cdots v_{n}\right)  $, the resulting Jacobian
is the product of the two Jacobians:%
\[
\frac{\partial(v_{1},v_{2},\cdots v_{n})}{\partial(u_{1},u_{2},\cdots u_{n}%
)}\cdot\frac{\partial(u_{1},u_{2},\cdots u_{n})}{\partial(x_{1},x_{2},\cdots
x_{n})}=\frac{\partial(v_{1},v_{2},\cdots v_{n})}{\partial(x_{1},x_{2},\cdots
x_{n})}.
\]

The definition of a Jacobian can lead to some interesting permutation rules as
the following examples\ illustrate. Consider a second order Jacobian
$\partial(u_{1},u_{2})/\partial(x_{1},x_{2})=\left(  \partial u_{1}/\partial
x_{1}\right)  \left(  \partial u_{2}/\partial x_{2}\right)  -\left(  \partial
u_{1}/\partial x_{2}\right)  \left(  \partial u_{2}/\partial x_{1}\right)  $,
which can be rearranged as%
\[
\frac{\partial(u_{1},u_{2})}{\partial(x_{1},x_{2})}\frac{\partial(x_{1}%
,x_{2})}{\partial(x_{1},x_{2})}+\frac{\partial(u_{2},x_{1})}{\partial
(x_{1},x_{2})}\frac{\partial(u_{1},x_{2})}{\partial(x_{1},x_{2})}%
+\frac{\partial(x_{1},u_{1})}{\partial(x_{1},x_{2})}\frac{\partial(u_{2}%
,x_{2})}{\partial(x_{1},x_{2})}=0.
\]
This can be symbolically written as
\begin{equation}
\partial(u_{1},u_{2})(x_{1},x_{2})+\partial(u_{2},x_{1})\partial(u_{1}%
,x_{2})+\partial(x_{1},u_{1})\partial(u_{2},x_{2})=0
\label{Permutation_Property}%
\end{equation}
by suppressing the common denominator in each term. The result expresses the
cyclic permutation of $u_{1},u_{2},x_{1}$ in the three terms with the
remaining variable $x_{2}$ in the same place in all terms. As a second
example, consider some quantity $u$ as a function of three variables $x,y,$
and $z$ and consider the following relation between the partial derivatives:%
\begin{equation}
\left(  \frac{\partial u}{\partial x}\right)  _{y}=\left(  \frac{\partial
u}{\partial x}\right)  _{y,z}+\left(  \frac{\partial u}{\partial z}\right)
_{x,y}\left(  \frac{\partial z}{\partial x}\right)  _{y}.
\label{Partial_Derivatives_Relation}%
\end{equation}
In terms of Jacobians, it can be written as%
\begin{equation}
\frac{\partial(u,y)}{\partial(x,y)}=\frac{\partial(u,y,z)}{\partial
(x,y,z)}+\frac{\partial(u,x,y)}{\partial(z,x,y)}\frac{\partial(z,y)}%
{\partial(x,y)}, \label{Partial_Derivatives_Relation_1}%
\end{equation}
which simplifies to
\begin{equation}
\partial(x,y,z)\partial(u,y)=\partial(y,z,u)\partial(x,y)+\partial
(z,u,x)\partial(y,y)+\partial(u,x,y)\partial(z,y),
\label{Permutation_Property_1}%
\end{equation}
where we have added a vanishing second term on the right because
$\partial(y,y)=0$. This relation is easily constructed by considering the
cyclic permutation of
\[
x,y,z,u
\]
by taking three consecutive terms at a time for the $3$-Jacobians, with the
remaining variable yielding the $2$-Jacobians in which the second entry is the
variable $y$, the variable that is held fixed in all derivatives in Eq.
(\ref{Partial_Derivatives_Relation}). The ordering $x,y,z$ in $x,y,z,u$ is
determined by the denominator $3$-Jacobian in the first term on the right in
Eq. (\ref{Partial_Derivatives_Relation_1}). By writing all the $3$-Jacobians
in the non-vanishing terms in Eq. (\ref{Permutation_Property_1}) so that $y$
is the second entry, and then suppressing the second entry, we obtain the
following relation%
\[
\partial(x,z)\partial(u,y)+\partial(z,u)\partial(x,y)+\partial(u,x)\partial
(z,y)=0,
\]
which is identical to the relation in Eq. (\ref{Permutation_Property}) if we
identify $u_{1}$ with $x$, $u_{2}$ with $z$, $x_{1}$ with $u$ and $x_{2}$ with
$y$.

We will use the Jacobians and their properties to first re-express the Maxwell
relations as follows%
\begin{align}
\frac{\partial(T_{0},S,N)}{\partial(V,S,N)}  &  =\frac{\partial(P_{0}%
,V,N)}{\partial(V,S,N)},\ \ \ \frac{\partial(T_{0},S,N)}{\partial(P_{0}%
,S,N)}=\frac{\partial(P_{0},V,N)}{\partial(P_{0},S,N)},\nonumber\\
\frac{\partial(P_{0},V,N)}{\partial(T_{0},V,N)}  &  =\frac{\partial
(T_{0},S,N)}{\partial(T_{0},V,N)},\ \ \ \frac{\partial(T_{0},S,N)}%
{\partial(P_{0},T_{0},N)}=\frac{\partial(P_{0},V,N)}{\partial(P_{0},T_{0},N)}.
\label{Maxwell_Jacobians}%
\end{align}

We now see a very important consequence of the use of the Jacobians. All four
Maxwell relations use the same numerators $\partial(T_{0},S,N)$ and
$\partial(P_{0},V,N)$. They use different denominators. Thus, they can all be
combined into one compact relation that can be simply written as
\begin{equation}
\partial(T_{0},S,N)\equiv\partial(P_{0},V,N). \label{Maxwell_Compact}%
\end{equation}
Here, the relation only has a meaning if each side is divided by one of the
possible denominators $\partial(V,S,N),\partial(P_{0},S,N),\partial
(T_{0},V,N)$ and $\partial(P_{0},T_{0},N)$ on both sides.

\subsection{Considerations in a Subspace\label{Marker_Subspace}}

It is very common to consider a function $F(x,y,z)$ in a subspace consisting
of $x,y$. This requires manipulating a $3$-Jacobians to construct a
$2$-Jacobians. of its argument. Thus, we may consider the $2$-Jacobian%
\[
\frac{\partial(F,y)}{\partial(x,y)},
\]
even though $F$ also depends on $z$. We can manipulate such Jacobians in the
normal way. For example, we can express it as
\begin{equation}
\left(  \frac{\partial F}{\partial x}\right)  _{y}=\frac{\partial
(F,y)}{\partial(x,y)}=-\frac{\partial(F,y)}{\partial(K,x)}\frac{\partial
(x,K)}{\partial(x,y)}=-\left(  \frac{\partial K}{\partial y}\right)  _{x}%
\frac{\partial(F,y)}{\partial(K,x)}, \label{Subspace_Reduction}%
\end{equation}
where $K(x,y,z)$ is another function. The derivation is tedious and has been
supplied in the Appendix. The situation can be generalized to many variables
$z_{1},z_{2},\cdots$ without much complications. We will not do this here.$\,$

\subsection{Some Transformation Rules \label{Marker_transformation rules}}

Let us consider a derivative of some quantity $R$ either with respect to $T$
or $P$ in case A below or at fixed $T$ or $P$ in case B below, which we wish
to express as a derivative involving $T_{0},P_{0}$ that are manipulated by the observer.

\begin{enumerate}
\item[A.] The derivative is at fixed $\mathbf{U}$, where $\mathbf{U}$ has any
two different elements from $E,V,S,\xi,P_{0}$ and $T_{0}.$
\end{enumerate}

We write the derivative as%
\begin{equation}
\left(  \frac{\partial R}{\partial T}\right)  _{U_{1},U_{2}}\equiv
\frac{\partial(R,U_{1},U_{2})}{\partial(T,U_{1},U_{2})}=\frac{\partial
(R,U_{1},U_{2})}{\partial(T_{0},U_{1},U_{2})}\frac{\partial(T_{0},U_{1}%
,U_{2})}{\partial(T,U_{1},U_{2})}=\left(  \frac{\partial R}{\partial T_{0}%
}\right)  _{U_{1},U_{2}}/\left(  \frac{\partial T}{\partial T_{0}}\right)
_{U_{1},U_{2}}. \label{TR_TP0}%
\end{equation}
Similarly, we have
\begin{equation}
\left(  \frac{\partial R}{\partial P}\right)  _{U_{1},U_{2}}\equiv
\frac{\partial(R,U_{1},U_{2})}{\partial(P,U_{1},U_{2})}=\frac{\partial
(R,U_{1},U_{2})}{\partial(P_{0},U_{1},U_{2})}\frac{\partial(P_{0},U_{1}%
,U_{2})}{\partial(P,U_{1},U_{2})}=\left(  \frac{\partial R}{\partial P_{0}%
}\right)  _{U_{1},U_{2}}/\left(  \frac{\partial P}{\partial P_{0}}\right)
_{U_{1},U_{2}}. \label{TR_PT0}%
\end{equation}

\begin{enumerate}
\item[B.] Let us consider a derivative with respect to $T_{0}$ at fixed
$U_{2}=P~$or $T$ ($U_{20}=P_{0}~$or $T_{0}$, as the case may be), but $U_{1}$
is any element from $E,V,S,\xi,P_{0}$ and $T_{0}$:
\begin{equation}
\left(  \frac{\partial R}{\partial T_{0}}\right)  _{U_{1},U_{2}}\equiv
\frac{\partial(R,U_{1},U_{2})}{\partial(T_{0},U_{1},U_{20})}\frac
{\partial(T_{0},U_{1},U_{20})}{\partial(T_{0},U_{1},U_{2})}=\frac
{\partial(R,U_{2},U_{1})}{\partial(T_{0},U_{20},U_{1})}/\left(  \frac{\partial
U_{2}}{\partial U_{20}}\right)  _{U_{1},U_{2}}. \label{TR_T0P}%
\end{equation}
Similarly,
\begin{equation}
\left(  \frac{\partial R}{\partial P_{0}}\right)  _{U_{1},U_{2}}\equiv
\frac{\partial(R,U_{1},U_{2})}{\partial(P_{0},U_{1},U_{20})}\frac
{\partial(P_{0},U_{1},U_{20})}{\partial(P_{0},U_{1},U_{2})}=\frac
{\partial(R,U_{2},U_{1})}{\partial(P_{0},U_{20},U_{1})}/\left(  \frac{\partial
U_{2}}{\partial U_{20}}\right)  _{U_{1},U_{2}}. \label{TR_P0T}%
\end{equation}
Let us now consider a derivative with respect to $T$ at fixed $P$ or with
respect to $P$ at fixed $T$; the derivative is at fixed $U_{1}$, where $U_{1}$
is any element from $E,V,S,\xi,P_{0}$ and $T_{0}.$%
\begin{equation}
\left(  \frac{\partial R}{\partial T}\right)  _{U_{1},P}\equiv\frac
{\partial(R,U_{1},P)}{\partial(T,U_{1},P)}=\frac{\partial(R,U_{1},P)}%
{\partial(T_{0},U_{1},P_{0})}/\frac{\partial(T,U_{1},P)}{\partial(T_{0}%
,U_{1},P_{0})}. \label{TR_TP}%
\end{equation}
Similarly,%
\begin{equation}
\left(  \frac{\partial R}{\partial P}\right)  _{U_{1},T}\equiv\frac
{\partial(R,U_{1},T)}{\partial(P,U_{1},T)}=\frac{\partial((R,U_{1}%
,T)}{\partial(T_{0},U_{1},P_{0})}/\frac{\partial(P,U_{1},T)}{\partial
(T_{0},U_{1},P_{0})}. \label{TR_PT}%
\end{equation}

\end{enumerate}

\section{Thermodynamic Potentials and Differentials\label{Marker_Potentials}}

\subsection{Equilibrium}

The forms of most useful thermodynamic potentials such as the enthalpy $H$,
the Helmholtz free energy $F$, and the Gibbs free energy $G$ of a system
$\Sigma$ in equilibrium are well known and are given in terms of the energy
$E(S,V,N)$ as
\begin{equation}
H=E+P_{0}V,\ F=E-T_{0}S,\ G=E-T_{0}S+P_{0}V, \label{Thermodynamic_Potentials}%
\end{equation}
where $T_{0},P_{0}$ are the temperature and pressure of the system; they are
also the temperature and/or pressure of the medium, depending on the medium
$\widetilde{\Sigma}$. Here, we are considering a system with fixed number of
particles. For the enthalpy, the medium $\widetilde{\Sigma}(P_{0})$ containing
the system exerts a fixed pressure $P_{0}$. For the Helmholtz free energy, the
medium $\widetilde{\Sigma}(T_{0})$ containing the system creates a fixed
temperature $T_{0}$. For the Gibbs free energy, the medium $\widetilde{\Sigma
}(T_{0},P_{0})$ containing the system exerts a fixed pressure $P_{0}$ and
creates a fixed temperature $T_{0}$. The potentials are Legendre transforms in
that the potentials are functions of the fields ($T_{0},P_{0}$) rather than
the observables ($E,V$) as the case may be. These potentials have the desired
property that they attain their minimum when the system is in equilibrium, as
discussed in I.

\subsection{Internal Equilibrium}

When the system is in internal equilibrium, we find from the Gibbs fundamental
relation for fixed $N$, which is obtained from setting $dN=0$ in Eq.
(\ref{Gibbs_Fundamental_relation_Internal}):
\begin{equation}
dE=TdS-PdV-Ad\xi, \label{Energy_relation_Internal/n}%
\end{equation}
where we have also introduced a single internal variable $\xi$ to allow us to
discuss non-equilibrium systems that are not in equilibrium with their medium
but are in internal equilibrium.\ The consideration of many internal variables
is to simply replace
\[
Ad\xi\rightarrow\mathbf{A\cdot}d\mathbf{\xi},
\]
and will not cause any extra complication. Thus, we will mostly consider a
single internal variable, but the extension to many internal variables is trivial.

We are no longer going to exhibit the time-dependence in these variables for
the sake of notational simplicity of. Let us return to Eq.
(\ref{Energy_relation_Internal/n}). It should be compared with Eq.
(\ref{Gibbs_Fundamental_relation_Equilibrium}) which contains $T_{0},P_{0}$.
We rewrite Eq. (\ref{Energy_relation_Internal/n}) to show the non-equilibrium
contribution explicitly:%
\begin{equation}
dE=T_{0}dS-P_{0}dV+(T-T_{0})dS-(P-P_{0})dV-Ad\xi.
\label{Energy_relation_Internal_Equilibrium/n}%
\end{equation}
The last two terms are due to the non-equilibrium nature of the system in
internal equilibrium. It is now easy to see that%
\begin{align}
dH  &  =T_{0}dS+VdP_{0}+(T-T_{0})dS-(P-P_{0})dV-Ad\xi,\nonumber\\
dF  &  =-SdT_{0}-P_{0}dV+(T-T_{0})dS-(P-P_{0})dV-Ad\xi
,\label{Thermodynamic_Differential_Internal/n}\\
dG  &  =-SdT_{0}+VdP_{0}+(T-T_{0})dS-(P-P_{0})dV-Ad\xi.\nonumber
\end{align}
These potentials correspond to $\xi$ as an independent variable of the
potential. One can make a transformation of these potentials to potentials in
which the conjugate field $A_{0}$ of the medium is the independent variable by
adding $A_{0}\xi$. The resulting potentials will be denoted by a superscript A
on the potential:%
\[
E^{\text{A}}=E+A_{0}\xi,H^{\text{A}}=H+A_{0}\xi,F^{\text{A}}=F+A_{0}%
\xi,G^{\text{A}}=G+A_{0}\xi.
\]
However, as discussed in II, $A_{0}=0$. Thus, there is no difference in the
values of the two potentials and the transformation is of no use. In
equilibrium, the internal fields $T,P$ attain their equilibrium values
$T_{0},P_{0}$ of the medium, and the affinity $A$ vanishes identically because
of $A_{0}=0$.

\section{Maxwell Relations For Systems in Internal
Equilibrium\label{Marker_Maxwell_Relations}}

From now on, we will always consider the case of a constant $N$. Therefore, we
will no longer exhibit it anymore. The Maxwell relation in Eq.
(\ref{Maxwell_Compact}) will then be denoted simply as $\partial
(T_{0},S)\equiv\partial(P_{0},V).$ The field parameters that appear in the
Maxwell relation are the parameters $T_{0},P_{0}$ of the medium, which because
of the existence of equilibrium also represent the field parameters of the
system. The Maxwell relation is a relation between the pairs $T_{0},S$ and
$P_{0},V$, each pair formed by the extensive variable and its conjugate field.
We will call these pairs conjugate pairs in this work. For a system described
by only two conjugate pairs, there is only one possible Maxwell relation. For
a system described by three conjugate pairs, there will be three different
Maxwell relations between them. For a system described by $k$ conjugate pairs,
there will be $k(k-1)/2$ different Maxwell relations.

As the system in internal equilibrium is very similar in many respects with an
equilibrium system as discussed in I and II, there may be analogs of the
Maxwell relations for systems in internal equilibrium. The question then
arises as to the field parameters that must appear in the Maxwell relations
when the system is not in equilibrium, but only in internal equilibrium. We
now turn to answer this question. Because of the absence of equilibrium, we
must now also include the internal variable $\xi$ in the discussion. Thus, we
expect three different Maxwell relations between

\subsection{Maxwell relation $\partial(T,S,\xi)\equiv\partial(P,V,\xi)$ at
fixed $\xi$}

We start with Eq. (\ref{Energy_relation_Internal/n}) and observe that%

\begin{equation}
\left(  \frac{\partial E}{\partial S}\right)  _{V,\xi}=T,\left(
\frac{\partial E}{\partial V}\right)  _{S,\xi}=-P,\left(  \frac{\partial
E}{\partial\xi}\right)  _{S,V}=-A. \label{Energy_Fields/n}%
\end{equation}
Using the first two derivative at fixed $\xi$, we find that%

\[
\left(  \frac{\partial^{2}E}{\partial V\partial S}\bigskip\right)  _{\xi
}=\left(  \frac{\partial T}{\partial V}\right)  _{S,\xi},\ \left(
\frac{\partial^{2}E}{\partial S\partial V}\bigskip\right)  _{\xi}=-\left(
\frac{\partial P}{\partial S}\right)  _{V,\xi}.
\]
As we are allowed to interchange the order of derivatives in the above cross
derivative, we have%

\[
\left(  \frac{\partial T}{\partial V}\right)  _{S,\xi}=-\left(  \frac{\partial
P}{\partial S}\right)  _{V,\xi},
\]
which can be written using Jacobians as
\[
\frac{\partial(T,S,\xi)}{\partial(S,V,\xi)}=\frac{\partial(P,V,\xi)}%
{\partial(S,V,\xi)}.
\]
This suggests the existence of the Maxwell relation $\partial(T,S,\xi
)=\partial(P,V,\xi)$ between the conjugate pairs $T,S$ and $P,V$ at fixed
$\xi$. To check its validity for other potentials with $\xi$ as an independent
variable, we consider the differential $dG$ in Eq.
(\ref{Thermodynamic_Differential_Internal/n}) and note that
\begin{align*}
\left(  \frac{\partial G}{\partial T_{0}}\right)  _{P_{0},\xi}  &
=-S+(T-T_{0})\left(  \frac{\partial S}{\partial T_{0}}\right)  _{P_{0},\xi
}+(P_{0}-P)\left(  \frac{\partial V}{\partial T_{0}}\right)  _{P_{0},\xi},\\
\left(  \frac{\partial G}{\partial P_{0}}\right)  _{T_{0},\xi}  &
=V+(T-T_{0})\left(  \frac{\partial S}{\partial P_{0}}\right)  _{T_{0},\xi
}+(P_{0}-P)\left(  \frac{\partial V}{\partial T_{0}}\right)  _{P_{0},\xi}.
\end{align*}
We use these derivatives to evaluate the cross derivative $\left(
\partial^{2}G/\partial P_{0}\partial T_{0}\right)  _{\xi}\bigskip$ to conclude
that
\begin{align*}
&  -\left(  \frac{\partial S}{\partial P_{0}}\right)  _{T_{0},\xi}%
+(T-T_{0})\left(  \frac{\partial^{2}S}{\partial T_{0}\partial P_{0}}\right)
_{\xi}+\left(  \frac{\partial T}{\partial P_{0}}\right)  _{T_{0},\xi}\left(
\frac{\partial S}{\partial T_{0}}\right)  _{P_{0},\xi}\\
&  +(P_{0}-P)\left(  \frac{\partial^{2}V}{\partial P_{0}\partial T_{0}%
}\right)  _{\xi}+\left(  \frac{\partial V}{\partial T_{0}}\right)  _{P_{0}%
,\xi}-\left(  \frac{\partial P}{\partial P_{0}}\right)  _{T_{0},\xi}\left(
\frac{\partial V}{\partial T_{0}}\right)  _{P_{0},\xi}\\
&  =\left(  \frac{\partial V}{\partial T_{0}}\right)  _{P_{0},\xi}%
+(T-T_{0})\left(  \frac{\partial^{2}S}{\partial P_{0}\partial T_{0}}\right)
_{\xi}+\left(  \frac{\partial T}{\partial T_{0}}\right)  _{P_{0},\xi}\left(
\frac{\partial S}{\partial P_{0}}\right)  _{T_{0},\xi}\\
&  +(P_{0}-P)\left(  \frac{\partial^{2}V}{\partial P_{0}\partial T_{0}%
}\right)  _{\xi}-\left(  \frac{\partial S}{\partial P_{0}}\right)  _{T_{0}%
,\xi}-\left(  \frac{\partial P}{\partial T_{0}}\right)  _{P_{0},\xi}\left(
\frac{\partial V}{\partial P_{0}}\right)  _{T_{0},\xi}.
\end{align*}
This is simplified to yield%
\[
\left(  \frac{\partial T}{\partial P_{0}}\right)  _{T_{0},\xi}\left(
\frac{\partial S}{\partial T_{0}}\right)  _{P_{0},\xi}-\left(  \frac{\partial
T}{\partial T_{0}}\right)  _{P_{0},\xi}\left(  \frac{\partial S}{\partial
P_{0}}\right)  _{T_{0},\xi}=\left(  \frac{\partial P}{\partial P_{0}}\right)
_{T_{0},\xi}\left(  \frac{\partial V}{\partial T_{0}}\right)  _{P_{0},\xi
}-\left(  \frac{\partial P}{\partial T_{0}}\right)  _{T_{0},\xi}\left(
\frac{\partial V}{\partial P_{0}}\right)  _{P_{0},\xi}.
\]
In terms of the Jacobians, this can be written as%

\[
\frac{\partial(T,S,\xi)}{\partial(T_{0},P_{0},\xi)}=\frac{\partial(P,V,\xi
)}{\partial(T_{0},P_{0},\xi)},
\]
thus justifying the Maxwell relation%
\begin{equation}
\partial(T,S,\xi)=\partial(P,V,\xi) \label{Maxwell_Relation_TSPV/N}%
\end{equation}
at fixed $\xi$. This relation must be satisfied at every point on the curve
ABDE that describes the vitrification process. This Maxwell relation turns
into the identity

\bigskip%
\[
\frac{\partial(T,S,\xi)}{\partial(P_{0},S,\xi)}=\frac{\partial(P,V,\xi
)}{\partial(P_{0},S,\xi)}%
\]
for the enthalpy and
\begin{equation}
\frac{\partial(T,S,\xi)}{\partial(T_{0},V,\xi)}=\frac{\partial(P,V,\xi
)}{\partial(T_{0},V,\xi)} \label{Maxwell_Helmholtz/n}%
\end{equation}
for the Helmholtz free energy, and are easily verified.

\subsection{Maxwell relation $\partial(T,S,V)\equiv\partial(A,\xi,V)$ at fixed
$V$}

We again start with Eqs. (\ref{Energy_relation_Internal/n}) and
(\ref{Energy_Fields/n}) , and evaluate the cross derivative $\left(
\partial^{2}E/\partial S\partial\xi\right)  _{V}$ to obtain%
\[
\left(  \frac{\partial^{2}E}{\partial\xi\partial S}\bigskip\right)
_{V}=\left(  \frac{\partial T}{\partial\xi}\right)  _{S,V},\left(
\frac{\partial^{2}E}{\partial S\partial\xi}\bigskip\right)  _{V}=-\left(
\frac{\partial A}{\partial S}\right)  _{V,\xi}.
\]
We thus have%

\[
\left(  \frac{\partial T}{\partial\xi}\right)  _{S,V}=-\left(  \frac{\partial
A}{\partial S}\right)  _{V,\xi},
\]
which can be written using Jacobians as
\[
\frac{\partial(T,S,V)}{\partial(\xi,S,V)}=\frac{\partial(A,\xi,V)}%
{\partial(\xi,S,V)}.
\]
This suggests the existence of the Maxwell relation $\partial(T,S,V)=\partial
(A,\xi,V)$ between the conjugate pairs $T,S$ and $A,\xi$ $\ $at fixed $V$. To
check its validity for other potentials with $V$ as an independent variable,
we consider the differential $dF$ in Eq.
(\ref{Thermodynamic_Differential_Internal/n}) and note that%
\begin{align*}
\left(  \frac{\partial F}{\partial T_{0}}\right)  _{V,\xi}  &  =-S+(T-T_{0}%
)\left(  \frac{\partial S}{\partial T_{0}}\right)  _{V,\xi}\\
\left(  \frac{\partial F}{\partial\xi}\right)  _{V,T_{0}}  &  =-A+(T-T_{0}%
)\left(  \frac{\partial S}{\partial\xi}\right)  _{V,T_{0}}.
\end{align*}
We now evaluate the cross derivative $\left(  \partial^{2}F/\partial
\xi\partial T_{0}\right)  _{V}$ and obtain the equality%
\begin{align*}
&  -\left(  \frac{\partial S}{\partial\xi}\right)  _{T_{0},V}+\left(
\frac{\partial S}{\partial T_{0}}\right)  _{V,\xi}\left(  \frac{\partial
T}{\partial\xi}\right)  _{T_{0},V}+(T-T_{0})\left(  \frac{\partial^{2}%
S}{\partial\xi\partial T_{0}}\right)  _{V}\\
&  =-\left(  \frac{\partial A}{\partial T_{0}}\right)  _{V,\xi}+\left[
\left(  \frac{\partial T}{\partial T_{0}}\right)  _{V,\xi}-1\right]  \left(
\frac{\partial S}{\partial\xi}\right)  _{T_{0},V}+(T-T_{0})\left(
\frac{\partial^{2}S}{\partial\xi\partial T_{0}}\right)  _{V},
\end{align*}
which leads to the relation%
\[
\frac{\partial(A,\xi,V)}{\partial(T_{0},\xi,V)}=\frac{\partial(T,S,V)}%
{\partial(T_{0},\xi,V)}.
\]
This confirms that the Maxwell relation between the conjugate pairs $T,S$ and
$A,\xi$ $\ $at fixed $V$ is the following:%
\begin{equation}
\partial(T,S,V)=\partial(A,\xi,V). \label{Maxwell_Relation_TSA/N}%
\end{equation}

\subsection{Maxwell Relation $\partial(P,V,S)\equiv\partial(A,\xi,S)$ at fixed
$S$}

We again start with Eqs. (\ref{Energy_relation_Internal/n}) and
(\ref{Energy_Fields/n}), and evaluate the cross derivative $\left(
\partial^{2}E/\partial V\partial\xi\right)  _{S}$ to obtain%
\[
\left(  \frac{\partial^{2}E}{\partial\xi\partial V}\bigskip\right)
_{S}=-\left(  \frac{\partial P}{\partial\xi}\right)  _{S,V},\left(
\frac{\partial^{2}E}{\partial V\partial\xi}\bigskip\right)  _{S}=-\left(
\frac{\partial A}{\partial V}\right)  _{S,\xi}.
\]
We thus have%

\[
\left(  \frac{\partial P}{\partial\xi}\right)  _{S,V}=\left(  \frac{\partial
A}{\partial V}\right)  _{S,\xi},
\]
which can be written using Jacobians as
\[
\frac{\partial(P,V,S)}{\partial(\xi,V,S)}=-\frac{\partial(A,\xi,S)}%
{\partial(\xi,V,S)}.
\]
This suggests the existence of the Maxwell relation $\partial(P,V,S)=-\partial
(A,\xi,S)$ between the conjugate pairs $P,V$ and $A,\xi$ $\ $at fixed $S$. To
check its validity for other potentials with $S$ as an independent variable,
we consider the differential $dH$ in Eq.
(\ref{Thermodynamic_Differential_Internal/n}) and note that%
\begin{align*}
\left(  \frac{\partial H}{\partial P_{0}}\right)  _{S,\xi}  &  =V-(P-P_{0}%
)\left(  \frac{\partial V}{\partial P_{0}}\right)  _{S,\xi}\\
\left(  \frac{\partial F}{\partial\xi}\right)  _{S,P_{0}}  &  =-A-(P-P_{0}%
)\left(  \frac{\partial V}{\partial\xi}\right)  _{S,T_{0}}.
\end{align*}
We now evaluate the cross derivative $\left(  \partial^{2}H/\partial
\xi\partial P_{0}\right)  _{S}$ and obtain the equality%
\begin{align*}
&  \left(  \frac{\partial V}{\partial\xi}\right)  _{P_{0},S}-\left(
\frac{\partial V}{\partial P_{0}}\right)  _{S,\xi}\left(  \frac{\partial
P}{\partial\xi}\right)  _{P_{0},S}-(P-P_{0})\left(  \frac{\partial^{2}%
V}{\partial\xi\partial P_{0}}\right)  _{S}\\
&  =-\left(  \frac{\partial A}{\partial P_{0}}\right)  _{S,\xi}-\left[
\left(  \frac{\partial P}{\partial P_{0}}\right)  _{S,\xi}-1\right]  \left(
\frac{\partial V}{\partial\xi}\right)  _{P_{0},S}-(P-P_{0})\left(
\frac{\partial^{2}V}{\partial\xi\partial P_{0}}\right)  _{S},
\end{align*}
which leads to the relation%
\[
\frac{\partial(A,\xi,S)}{\partial(P_{0},\xi,S)}=-\frac{\partial(P,V,S)}%
{\partial(P_{0},\xi,S)}.
\]
This confirms that the Maxwell relation between the conjugate pairs $P,V$ and
$A,\xi$ $\ $at fixed $S$ is the following:%
\begin{equation}
\partial(P,V,S)=-\partial(A,\xi,S). \label{Maxwell_Relation_PVA/N}%
\end{equation}

One can easily check that this Maxwell relation also works with other
thermodynamic potentials like $F$ and $G$. We will satisfy ourselves by giving
the demonstration for $F$ only. The natural variables for $F$ are $T_{0},\xi$
and $V$; however, instead of using $V$ as the independent variable, we will
use $S$ as the independent variable so that it can be held fixed. For constant
$S$, the differential $dF$ from Eq.
(\ref{Thermodynamic_Differential_Internal/n}) reduces to
\[
\left.  dF\right\vert _{S}=-SdT_{0}-PdV-Ad\xi,
\]
so that
\begin{align*}
\left(  \frac{\partial F}{\partial T_{0}}\right)  _{S,\xi}  &  =-S-P\left(
\frac{\partial V}{\partial T_{0}}\right)  _{S,\xi}\\
\left(  \frac{\partial F}{\partial\xi}\right)  _{S,T_{0}}  &  =-A-P\left(
\frac{\partial V}{\partial\xi}\right)  _{S,T_{0}}.
\end{align*}
Now evaluating the cross derivative\ $\left(  \partial^{2}F/\partial
\xi\partial T_{0}\right)  _{S}$, we find that
\begin{align*}
&  -\left(  \frac{\partial A}{\partial T_{0}}\right)  _{S,\xi}-\left(
\frac{\partial P}{\partial T_{0}}\right)  _{S,\xi}\left(  \frac{\partial
V}{\partial\xi}\right)  _{T_{0},S}-P\left(  \frac{\partial^{2}V}{\partial
\xi\partial T_{0}}\right)  _{S}\\
&  =-\left(  \frac{\partial V}{\partial T_{0}}\right)  _{S,\xi}\left(
\frac{\partial P}{\partial\xi}\right)  _{T_{0},S}-P\left(  \frac{\partial
^{2}V}{\partial\xi\partial P_{0}}\right)  _{S}.
\end{align*}
This now immediately leads to
\begin{equation}
\frac{\partial(A,\xi,S)}{\partial(T_{0},\xi,S)}=-\frac{\partial(P,V,S)}%
{\partial(T_{0},\xi,S)}, \label{Maxwell_Relation_F/S/n}%
\end{equation}
and confirms our claim that the Maxwell relation is given by Eq.
(\ref{Maxwell_Relation_PVA/N}).

\subsection{General Maxwell Relations with system variables only}

We wish to emphasize that the Maxwell relation in
Eq.\ (\ref{Maxwell_Relation_F/S/n}) requires keeping $S$ fixed so that we must
divide Eq. (\ref{Maxwell_Relation_PVA/N}) by $\partial(T_{0},\xi,S)$ on both
sides. We must not use the independent variables $T_{0},\xi$ and $V$ of $F$
for the division and keep $T_{0}$ fixed. This will not give be a Maxwell
relation. We demonstrate this explicitly by evaluating $\left(  \partial
^{2}F/\partial\xi\partial V\right)  _{T_{0}}$ two different ways and equating
the results. A simple calculation yields
\begin{align*}
&  -\left(  \frac{\partial P}{\partial\xi}\right)  _{V,T_{0}}+\left(
\frac{\partial S}{\partial V}\right)  _{T_{0},\xi}\left(  \frac{\partial
T}{\partial\xi}\right)  _{V,T_{0}}+(T-T_{0})\left(  \frac{\partial^{2}%
S}{\partial V\partial\xi}\right)  _{T_{0}}\\
&  =-\left(  \frac{\partial A}{\partial V}\right)  _{T_{0},\xi}+\left(
\frac{\partial S}{\partial\xi}\right)  _{V,T_{0}}\left(  \frac{\partial
T}{\partial V}\right)  _{T_{0},\xi}+(T-T_{0})\left(  \frac{\partial^{2}%
S}{\partial V\partial\xi}\right)  _{T_{0}}.
\end{align*}
In terms of Jacobians, the above equation can be rewritten as%
\begin{equation}
\ \ \frac{\partial(A,\xi,T_{0})}{\partial(V,\xi,T_{0})}=-\frac{\partial
(P,V,T_{0})}{\partial(V,\xi,T_{0})}+\frac{\partial(T,S,T_{0})}{\partial
(V,\xi,T_{0})}. \label{Maxwell_Relation_F/T0/n}%
\end{equation}
This relation from the cross derivative requires keeping $T_{0}$ fixed.
However, $T_{0}$ characterizes the medium and only indirectly characterizes
the system in internal equilibrium. In a similar way, using the cross
derivatives of the Gibbs free energy at fixed $T_{0}$, and at fixed $P_{0}$,
we find the following relations:%

\begin{equation}
\frac{\partial(A,\xi,P_{0})}{\partial(T_{0},P_{0,}\xi)}=\frac{\partial
(T,S,P_{0})}{\partial(T_{0},P_{0},\xi)}+\frac{\partial(P,V,P_{0})}%
{\partial(T_{0},P_{0},\xi)},\ \ \ \frac{\partial(A,\xi,T_{0})}{\partial
(T_{0},P_{0,}\xi)}=\frac{\partial(P,V,T_{0})}{\partial(T_{0},P_{0,}\xi)}%
-\frac{\partial(T,S,T_{0})}{\partial(T_{0},P_{0,}\xi)}.
\label{Maxwell_Relation_G/T0_P0/n}%
\end{equation}
We now wish to observe that the Maxwell relations appear only when we keep the
quantities of the system $T,P,S,V,A,$ or $\xi$ fixed. We have already seen the
Maxwell relations with fixed $S,V$, and $\xi$. We will now consider keeping
$T$ fixed to demonstrate our point. For fixed $T$, we obtain the following
Maxwell relation%
\begin{equation}
\frac{\partial(A,\xi,T)}{\partial(T_{0},\xi,T)}=-\frac{\partial(P,V,T)}%
{\partial(T_{0},\xi,T)}, \label{Maxwell_Relation_F/T/n}%
\end{equation}
as can easily be checked by evaluating the cross derivative $\left(
\partial^{2}F/\partial\xi\partial T_{0}\right)  _{T}$ at fixed $T$. The
calculation is identical to that carried out in obtaining Eq.
(\ref{Maxwell_Relation_F/S/n}). One can easily check that keeping $P$ or $A$
also gives us new Maxwell relations%
\[
\frac{\partial(A,\xi,P)}{\partial(T_{0},\xi,P)}=\frac{\partial(T,S,P)}%
{\partial(T_{0},\xi,P)},\frac{\partial(T,S,A)}{\partial(\xi,T_{0},A)}%
=-\frac{\partial(P,V,A)}{\partial(T_{0},\xi,A)}.
\]

\section{Clausius-Clapeyron Relation\label{Marker_Clausius_Clapeyron_Relation}%
}

As a system in internal equilibrium is not very different from that in
equilibrium, except that its Gibbs free energy $G(t)$ continuously decreases
until it reaches equilibrium with the medium, it is possible for the system to
exist in two distinct phases that have the same Gibbs free energy at some
instant. Such a non-equilibrium phase transition situation will arise, for
example, when an isotropic supercooled liquid can turn into a liquid crystal
phase. This is not a novel idea as there are several attempts in the
literature \cite[and references therin]{Onuki,Sugar,Allahverdyan,Arndt} where
such non-equilibrium phase transitions have been investigated. Therefore, let
us now consider the possibility of the system being in two different phases at
some time. As experiments are carried out by controlling observables only and
not the internal variables, it is important to consider thermodynamic
quantities as a function of $\mathbf{X~}$only, and not of $\mathbf{X,I}$ in
all cases. Restricting ourselves to a single internal variable $\xi$, and to
$E$ and $V$, we will treat thermodynamic quantities not only as a function of
three independent variables, but will also have the need to consider them as a
function of observables or associated fields $T_{0},P_{0}$. In particular, the
Clausius-Clapeyron relation is obtained in the $T_{0}$-$P_{0}$ plane, a
subspace; see Sect. \ref{Marker_Subspace}.

Let us consider the two phases, which we denote by $1$ and $2$, in the system.
We will use subscripts $1$ and $2$ to refer to the quantities in the two
phases. In internal equilibrium, the entropy $S$ of the system is a function
of the averages $\mathbf{X(}t)\mathbf{,I(}t\mathbf{)}$ along with the fixed
number of particles $N$. It is important to include $N$ in our consideration
as the two phases will contain number of particles $N_{1}$ and $N_{2}$ that
are not constant, except in equilibrium. Obviously,%
\[
\mathbf{X(}t)=\mathbf{X}_{1}\mathbf{(}t)+\mathbf{X}_{2}\mathbf{(}%
t),\mathbf{I(}t)=\mathbf{I}_{1}\mathbf{(}t)+\mathbf{I}_{2}\mathbf{(}%
t),N=N_{1}(t)+N_{2}(t).
\]
Then, we can express the entropy of the system as a sum over the two phases:%
\[
S(\mathbf{X(}t),\mathbf{I(}t),N)=S_{1}(\mathbf{X}_{1}\mathbf{(}t),\mathbf{I}%
_{1}\mathbf{(}t),N_{1}(t))+S_{2}(\mathbf{X}_{2}\mathbf{(}t),\mathbf{I}%
_{2}\mathbf{(}t),N_{2}(t)),
\]
which takes its maximum possible value for given $\mathbf{X(}t),\mathbf{I(}%
t),N$ in internal equilibrium. Thus,%
\[
dS(\mathbf{X(}t),\mathbf{I(}t),N)=dS_{1}(\mathbf{X}_{1}\mathbf{(}%
t),\mathbf{I}_{1}\mathbf{(}t),N_{1}(t))+dS_{2}(\mathbf{X}_{2}\mathbf{(}%
t),\mathbf{I}_{2}\mathbf{(}t),N_{2}(t))=0
\]
in internal equilibrium. This can only happen if
\[
\mathbf{y}_{1}(t)=\mathbf{y}_{2}(t),\ \ \mathbf{a}_{1}\mathbf{(}%
t)\ \mathbf{=a}_{2}\mathbf{(}t\mathbf{),\ \ }\ \mu_{1}(t)/T_{1}(t)=\ \mu
_{2}(t)/T_{2}(t);
\]
see Eqs. (\ref{Gibbs_Fundamental_relation_0}), (\ref{Field_Variables_1}) and
(\ref{Field_Variables_0}).

For the restricted case under consideration, this results in the equality%
\[
T_{1}(t)=T_{2}(t),\ \ P_{1}\mathbf{(}t)\ \mathbf{=}P_{2}\mathbf{(}%
t\mathbf{),\ \ }\ \mu_{1}(t)=\ \mu_{2}(t),\ \ A_{1}(t)=A_{2}(t)
\]
for the internal fields and affinity along the coexistence of the two phases.
It also follows from the continuity of the Gibbs free energy
\cite{Gujrati-Non-Equilibrium-I} that the Gibbs free energies of the two pure
phases ($N_{1}=N$ and $N_{2}=N$) must be equal at the coexistence. We will
only consider the two pure phases below, and not a mixture of the two. As the
numbers of particles in the two pure phases are constant, we will no longer
consider them anymore in the discussion.

We now consider the $T_{0}$-$P_{0}$ plane, relevant for the observation of
coexistence. Since the Gibbs free energy is continuous along the transition line,%

\[
\Delta G(T_{0},P_{0}(T_{0}))=0
\]
where $P_{0}(T_{0})$ is the pressure along the transition line. Thus,%

\[
d\Delta G=\Delta\left(  \frac{\partial G}{\partial T_{0}}\right)  _{P_{0}%
}dT_{0}+\Delta\left(  \frac{\partial G}{\partial P_{0}}\right)  _{T_{0}}%
dP_{0}.
\]
Using $d\Delta G=0$ yields%

\begin{equation}
\left.  \frac{dT_{0}}{dP_{0}}\right\vert _{\text{coex}}\Delta\left(
\frac{\partial G}{\partial T_{0}}\right)  _{P_{0}}+\Delta\left(
\frac{\partial G}{\partial P_{0}}\right)  _{T_{0}}=0 \label{dG_Coexistence}%
\end{equation}
along the coexistence. Using $dG$ from Eq.
(\ref{Thermodynamic_Differential_Internal/n}) gives us%

\begin{equation}
\Delta\left(  \frac{\partial G}{\partial T_{0}}\right)  _{P_{0}}=-\Delta
S+(T-T_{0})\Delta\left(  \frac{\partial S}{\partial T_{0}}\right)  _{P_{0}%
}-(P-P_{0})\Delta\left(  \frac{\partial V}{\partial T_{0}}\right)  _{P_{0}%
}-A\Delta\left(  \frac{\partial\xi}{\partial T_{0}}\right)  _{P_{0}}
\label{Delta_dG_T0}%
\end{equation}

\begin{equation}
\Delta\left(  \frac{\partial G}{\partial P_{0}}\right)  _{T_{0}}=\Delta
V+(T-T_{0})\Delta\left(  \frac{\partial S}{\partial P_{0}}\right)  _{T_{0}%
}-(P-P_{0})\Delta\left(  \frac{\partial V}{\partial P_{0}}\right)  _{T_{0}%
}-A\Delta\left(  \frac{\partial\xi}{\partial P_{0}}\right)  _{T_{0}}
\label{Delta_dG_P0}%
\end{equation}

Putting the above two equations in Eq. (\ref{dG_Coexistence}), we get the
following Clausius-Clapeyron equation for coexistence of phases in internal
equilibrium%
\begin{equation}
\left.  \frac{dT_{0}}{dP_{0}}\right\vert _{\text{coex}}=\frac{\Delta
V+(T-T_{0})\Delta\left(  \partial S/\partial P_{0}\right)  _{T_{0}}%
-(P-P_{0})\Delta\left(  \partial V/\partial P_{0}\right)  _{T_{0}}%
-A\Delta\left(  \partial\xi/\partial P_{0}\right)  _{T_{0}}}{\Delta
S-(T-T_{0})\Delta\left(  \partial S/\partial T_{0}\right)  _{P_{0}}%
+(P-P_{0})\Delta\left(  \partial V/\partial T_{0}\right)  _{P_{0}}%
+A\Delta\left(  \partial\xi/\partial T_{0}\right)  _{P_{0}}}.
\label{Clausius-Claperon_Eq/n}%
\end{equation}
We now express $\left(  \partial S/\partial P_{0}\right)  _{T_{0}}$ in terms
of $\left(  \partial V/\partial T_{0}\right)  _{P_{0}}$ by using the Maxwell
relation $\partial(P,V)=\partial(T,S)$ and by using Eq.
(\ref{Subspace_Reduction}) ($F\rightarrow S,K\rightarrow V,x\rightarrow
P_{0},$ and $y\rightarrow T_{0})$ as follows:%
\[
\frac{\partial(S,T_{0})}{\partial(P_{0},T_{0})}=-\frac{\partial(S,T_{0}%
)}{\partial(S,T)}\frac{\partial(P,V)}{\partial(P_{0},V)}\frac{\partial
(P_{0},V)}{\partial(P_{0},T_{0})},
\]
which immediately gives%

\begin{equation}
\left(  \frac{\partial S}{\partial P_{0}}\right)  _{T_{0}}=-\frac{\left(
\partial P/\partial P_{0}\right)  _{V}}{\left(  \partial T/\partial
T_{0}\right)  _{S}}\left(  \frac{\partial V}{\partial T_{0}}\right)  _{P_{0}},
\label{S_P0_V_T0_relation}%
\end{equation}
which can now be used in the Clausius-Clapeyron equation to express it in
terms of measurable quantities assuming that $P,T$ can be measured. In
equilibrium, $T=T_{0},P=P_{0}$ and $A=0$, so that the above equation reduces
to the well-known version%
\begin{equation}
\left.  \frac{dT_{0}}{dP_{0}}\right\vert _{\text{coex}}^{\text{(eq)}}%
=\frac{\Delta V}{\Delta S}, \label{Clausius-Claperon_Eq}%
\end{equation}
as expected.

\section{\textbf{Response functions in Internal Equilibrium}%
\label{Marker_Response_Functions}}

\subsection{$\bar{C}_{P}$\textbf{\ and }$\bar{C}_{V}$\textbf{ }}

The heat capacities with respect to the internal temperature at fixed $P$ or
$V$ are%
\begin{align*}
\bar{C}_{P,\xi}  &  =T\left(  \frac{\partial S}{\partial T}\right)  _{P,\xi
},\ \ \bar{C}_{V,\xi}=T\left(  \frac{\partial S}{\partial T}\right)  _{V,\xi
},\\
\bar{C}_{P}  &  =T\left(  \frac{\partial S}{\partial T}\right)  _{P}%
,\ \ \bar{C}_{V}=T\left(  \frac{\partial S}{\partial T}\right)  _{V}%
\end{align*}
We again start from the fundamental relation in Eq.
(\ref{Fundamental relation0}) and evaluate the derivative%

\[
T\left(  \frac{\partial S}{\partial T}\right)  _{P,\xi}=T\left(
\frac{\partial S}{\partial T}\right)  _{V,\xi}+T\left(  \frac{\partial
S}{\partial V}\right)  _{T,\xi}\left(  \frac{\partial V}{\partial T}\right)
_{P,\xi},
\]
which can be rewritten in two equivalent forms%
\begin{equation}
\overline{C}_{P,\xi}=\overline{C}_{V,\xi}+T\left[  \left(  \frac{\partial
S}{\partial P}\right)  _{T,\xi}\left(  \frac{\partial V}{\partial T}\right)
_{P,\xi}\right]  /\left(  \frac{\partial V}{\partial P}\right)  _{T,\xi}
\label{Internal_Heat_Capacities_Relation}%
\end{equation}
or%
\begin{equation}
\overline{C}_{P,\xi}=\overline{C}_{V,\xi}+T\left(  \frac{\partial P}{\partial
T}\right)  _{V,\xi}\left(  \frac{\partial V}{\partial T}\right)  _{P,\xi},
\label{Internal_Heat_Capacities_Relation_0}%
\end{equation}
where we have used the Maxwell relation in Eq. (\ref{Maxwell_Relation_TSPV/N})
after we divide it by $\partial(V,T,\xi)$. As $\left(  \partial S/\partial
P\right)  _{V,\xi}$ is not directly measurable, the identity in Eq.
(\ref{Internal_Heat_Capacities_Relation_0}) is more useful from a practical
point of view. However, we need to transform the various derivatives in it to
the derivatives with respect to $T_{0}$ at fixed $P_{0}$ or $V$ by using the
transformation rules in Sect. \ref{Marker_transformation rules}, as it is the
pair $T_{0},P_{0}$ that can be manipulated by the observer. However, the
identities still contains $\bar{C}_{P,\xi}$ and $\bar{C}_{V,\xi}$, which are
defined with respect to $T$, and not with respect to $T_{0}$. Therefore, we
now turn to heat capacities obtained as a derivative with respect to $T_{0}$.

\subsection{$C_{P}$\textbf{\ and }$C_{V}$\textbf{ }}

From Eq. (\ref{Heat_Transfer})$,$ we have
\begin{align}
C_{P}  &  \equiv\left(  \frac{\partial Q}{\partial T_{0}}\right)  _{P_{0}%
}\equiv T\left(  \frac{\partial S}{\partial T_{0}}\right)  _{P_{0}}%
,C_{V}\equiv\left(  \frac{\partial Q}{\partial T_{0}}\right)  _{V}\equiv
T\left(  \frac{\partial S}{\partial T_{0}}\right)  _{V}, \label{Heat_Capacity}%
\\
C_{P,\xi}  &  \equiv\left(  \frac{\partial Q}{\partial T_{0}}\right)
_{P_{0},\xi}\equiv T\left(  \frac{\partial S}{\partial T_{0}}\right)
_{P_{0},\xi},C_{V,\xi}\equiv\left(  \frac{\partial Q}{\partial T_{0}}\right)
_{V,\xi}\equiv T\left(  \frac{\partial S}{\partial T_{0}}\right)  _{V,\xi}.
\label{Heat_Capacity_Xi}%
\end{align}
It would have been more appropriate to express the capacities $C_{P}%
$\textbf{\ }and\textbf{ }$C_{P,\xi}$ as $C_{P_{0}}$\textbf{\ }and\textbf{
}$C_{P_{0},\xi}$, but we will use the simpler notation. This should cause no
confusion. Introducing the expansion coefficient
\begin{equation}
\alpha_{P}\equiv\frac{1}{V}\left(  \frac{\partial V}{\partial T_{0}}\right)
_{P_{0}},\alpha_{P,\xi}\equiv\frac{1}{V}\left(  \frac{\partial V}{\partial
T_{0}}\right)  _{P_{0},\xi} \label{Expansion_Coefficients}%
\end{equation}
we find that
\[
\frac{C_{P}}{\alpha_{P}}=TV\frac{\partial(S,P_{0})/\partial(T_{0},P_{0}%
)}{\partial(V,P_{0})/\partial(T_{0},P_{0})}=TV\frac{\partial(S,P_{0}%
)}{\partial(V,P_{0})}=TV\left(  \frac{\partial S}{\partial V}\right)  _{P_{0}%
}.
\]
The same discussion can be applied to $C_{P,\xi}$ and $\alpha_{P,\xi}$ with a
similar result%
\[
\frac{C_{P,\xi}}{\alpha_{P,\xi}}=TV\frac{\partial(S,P_{0},\xi)/\partial
(T_{0},P_{0},\xi)}{\partial(V,P_{0},\xi)/\partial(T_{0},P_{0},\xi)}%
=TV\frac{\partial(S,P_{0},\xi)}{\partial(V,P_{0},\xi)}=TV\left(
\frac{\partial S}{\partial V}\right)  _{P_{0},\xi}.
\]

Let us now consider the relation between $C_{P,\xi}$\textbf{\ }and\textbf{
}$C_{V,\xi}$ and between $C_{P}$\textbf{\ }and\textbf{ }$C_{V}$, for
which\textbf{ }we consider $S$ as a function of $T,V$ and $\xi$, which follows
from Eq. (\ref{Energy_relation_Internal/n}), so that
\begin{equation}
dS=\frac{\partial S}{\partial T}dT+\frac{\partial S}{\partial V}%
dV+\frac{\partial S}{\partial\xi}d\xi. \label{Fundamental relation0}%
\end{equation}
Therefore,
\[
\left(  \frac{\partial S}{\partial T_{0}}\right)  _{P_{0},\xi}=\left(
\frac{\partial S}{\partial T}\right)  _{V,\xi}\left(  \frac{\partial
T}{\partial T_{0}}\right)  _{P_{0},\xi}+\left(  \frac{\partial S}{\partial
V}\right)  _{T,\xi}\left(  \frac{\partial V}{\partial T_{0}}\right)
_{P_{0},\xi}.
\]
Now, using Eq. (\ref{TR_TP0}), we have
\begin{equation}
\left(  \frac{\partial S}{\partial T}\right)  _{V,\xi}=\left(  \frac{\partial
S}{\partial T_{0}}\right)  _{V,\xi}/\left(  \frac{\partial T}{\partial T_{0}%
}\right)  _{V,\xi} \label{S_T_V_derivative}%
\end{equation}
Similarly, using the Maxwell relation in Eq. (\ref{Maxwell_Helmholtz/n}) we
have
\begin{equation}
\left(  \frac{\partial S}{\partial V}\right)  _{T,\xi}=\frac{\partial
(V,P,\xi)}{\partial(V,T_{0},\xi)}\frac{\partial(V,T_{0},\xi)}{\partial
(V,T,\xi)}=\left(  \frac{\partial P}{\partial T_{0}}\right)  _{V,\xi}/\left(
\frac{\partial T}{\partial T_{0}}\right)  _{V,\xi}. \label{S_V_T_derivative}%
\end{equation}
We thus finally obtain%
\[
\left(  \frac{\partial S}{\partial T_{0}}\right)  _{P_{0},\xi}\left(
\frac{\partial T}{\partial T_{0}}\right)  _{V,\xi}=\left(  \frac{\partial
S}{\partial T_{0}}\right)  _{V,\xi}\left(  \frac{\partial T}{\partial T_{0}%
}\right)  _{P_{0},\xi}+\left(  \frac{\partial P}{\partial T_{0}}\right)
_{V,\xi}\left(  \frac{\partial V}{\partial T_{0}}\right)  _{P_{0},\xi},
\]
After multiplying by $T$ on both sides, we obtain the desired relation between
$C_{P,\xi}$ and $C_{V,\xi}$ for the non-equilibrium case%
\begin{equation}
C_{P,\xi}\left(  \frac{\partial T}{\partial T_{0}}\right)  _{V,\xi}=C_{V,\xi
}\left(  \frac{\partial T}{\partial T_{0}}\right)  _{P_{0},\xi}+T\left(
\frac{\partial P}{\partial T_{0}}\right)  _{V,\xi}\left(  \frac{\partial
V}{\partial T_{0}}\right)  _{P_{0},\xi}.
\label{Internal_Heat_Capacities_T0_Relation}%
\end{equation}
This relation generalize the following standard equilibrium relation:%
\[
C_{P}^{\text{eq}}=C_{V}^{\text{eq}}+T_{0}\left(  \frac{\partial P_{0}%
}{\partial T_{0}}\right)  _{V}\left(  \frac{\partial V}{\partial T_{0}%
}\right)  _{P_{0}},
\]
obtained by setting
\[
\left(  \frac{\partial T}{\partial T_{0}}\right)  _{V,\xi}=\left(
\frac{\partial T}{\partial T_{0}}\right)  _{P_{0},\xi}=1.
\]

We can obtain a standard form of the above heat capacity relations as in Eq.
(\ref{Internal_Heat_Capacities_Relation}) as follows:%

\[
C_{V,\xi}=T\frac{\partial(S,V,\xi)}{\partial(T_{0},V,\xi)}=T\frac
{\partial(S,V,\xi)/\partial(T_{0},P_{0},\xi)}{\partial(T_{0},V,\xi
)/\partial(T_{0},P_{0},\xi)}.
\]
We thus finally have
\begin{equation}
C_{P,\xi}=C_{V,\xi}+T\frac{(\partial S/\partial P_{0})_{T_{0},\xi}(\partial
V/\partial T_{0})_{P_{0},\xi}}{(\partial V/\partial P_{0})_{T_{0},\xi}%
}=C_{V,\xi}+T\left(  \frac{\partial S}{\partial V}\right)  _{T_{0},\xi}\left(
\frac{\partial V}{\partial T_{0}}\right)  _{P_{0},\xi},
\label{Internal_Heat_Capacities_T0_Relation00}%
\end{equation}
which is an extension of Eq. (\ref{Internal_Heat_Capacities_Relation}).
Although tedious, it is straightforward to show that this relation is
identical to the above relation. One needs to evaluate $\left(  \partial
S/\partial V\right)  _{T_{0},\xi}$ as follows:%
\begin{align*}
\left(  \frac{\partial S}{\partial V}\right)  _{T_{0},\xi}  &  =\frac
{\partial(S,T_{0},\xi)}{\partial(V,T,\xi)}\frac{\partial(V,T,\xi)}%
{\partial(V,T_{0},\xi)}\\
&  =\left(  \frac{\partial T}{\partial T_{0}}\right)  _{V,\xi}\left[  \left(
\frac{\partial S}{\partial V}\right)  _{T,\xi}\left(  \frac{\partial T_{0}%
}{\partial T}\right)  _{V,\xi}-\left(  \frac{\partial S}{\partial T}\right)
_{V,\xi}\left(  \frac{\partial T_{0}}{\partial V}\right)  _{T,\xi}\right]  ,
\end{align*}
where we must now use Eqs. (\ref{S_V_T_derivative}) and
(\ref{S_T_V_derivative}). We finally obtain%
\begin{equation}
\left(  \frac{\partial S}{\partial V}\right)  _{T_{0},\xi}=\left(
\frac{\partial T}{\partial T_{0}}\right)  _{V,\xi}\left[  \left(
\frac{\partial P}{\partial T_{0}}\right)  _{V,\xi}\left(  \frac{\partial
T_{0}}{\partial T}\right)  _{V,\xi}\left(  \frac{\partial T_{0}}{\partial
T}\right)  _{V,\xi}-\left(  \frac{\partial S}{\partial T_{0}}\right)  _{V,\xi
}\left(  \frac{\partial T_{0}}{\partial T}\right)  _{V,\xi}\left(
\frac{\partial T_{0}}{\partial V}\right)  _{T,\xi}\right]  .
\label{S_V_T0_derivative}%
\end{equation}
The equivalence is now established by the use of the permutation property
given in Eq. (\ref{Permutation_Property}). In a similar fashion, we find that
\begin{equation}
C_{P}=C_{V}+T\left(  \frac{\partial S}{\partial V}\right)  _{T_{0}}\left(
\frac{\partial V}{\partial T_{0}}\right)  _{P_{0}},
\label{CP_CV_relation_full}%
\end{equation}
where we must use, see Sect. \ref{Marker_Subspace},
\begin{equation}
\left(  \frac{\partial S}{\partial V}\right)  _{T_{0}}=\left(  \frac{\partial
T}{\partial T_{0}}\right)  _{V}\left[  \left(  \frac{\partial P}{\partial
T_{0}}\right)  _{V}\left(  \frac{\partial T_{0}}{\partial T}\right)
_{V}\left(  \frac{\partial T_{0}}{\partial T}\right)  _{V}-\left(
\frac{\partial S}{\partial T_{0}}\right)  _{V}\left(  \frac{\partial T_{0}%
}{\partial T}\right)  _{V}\left(  \frac{\partial T_{0}}{\partial V}\right)
_{T}\right]  \label{S_V_T0_only_derivative}%
\end{equation}
obtained in a similar fashion as Eq. (\ref{S_V_T0_derivative}).

It is important at this point to relate $C_{P}$ with $C_{P,\xi}$ and $C_{V}$
with $C_{V,\xi}.$ For this, it is convenient to consider the differential $dS$
by treating $S$ as a function of $T_{0},P_{0}$ and $\xi$. We find that
\begin{equation}
C_{P}=C_{P,\xi}+T\left(  \frac{\partial S}{\partial\xi}\right)  _{T_{0},P_{0}%
}\left(  \frac{\partial\xi}{\partial T_{0}}\right)  _{P_{0}},\ \ \ C_{V}%
=C_{V,\xi}+T\left(  \frac{\partial S}{\partial\xi}\right)  _{P_{0},V}\left(
\frac{\partial\xi}{\partial T_{0}}\right)  _{V}. \label{C_C_Xi_Relation}%
\end{equation}

\subsection{Compressibilities $K_{T}$ and $K_{S}$}

The two important isothermal compressibilities are%
\[
K_{T}\equiv-\frac{1}{V}\left(  \frac{\partial V}{\partial P_{0}}\right)
_{T_{0}},\ \ K_{T,\xi}\equiv-\frac{1}{V}\left(  \frac{\partial V}{\partial
P_{0}}\right)  _{T_{0},\xi},
\]
which we need to relate to the corresponding adiabatic compressibility%
\[
K_{S}\equiv-\frac{1}{V}\left(  \frac{\partial V}{\partial P_{0}}\right)
_{S},\ \ K_{S,\xi}\equiv-\frac{1}{V}\left(  \frac{\partial V}{\partial P_{0}%
}\right)  _{S,\xi}.
\]
However, we first consider the relation between the compressibility and the
expansion coefficient. We find that
\[
\frac{K_{T}}{\alpha_{P}}=\frac{\partial(V,T_{0})/\partial(T_{0},P_{0}%
)}{\partial(V,P_{0})/\partial(T_{0},P_{0})}=\frac{\partial(V,T_{0})}%
{\partial(V,P_{0})}=\left(  \frac{\partial T_{0}}{\partial P_{0}}\right)
_{V}.
\]
The same discussion can be applied to $K_{T,\xi}$ and $\alpha_{P,\xi}$ with a
similar result%
\[
\frac{K_{T,\xi}}{\alpha_{P,\xi}}=\left(  \frac{\partial T_{0}}{\partial P_{0}%
}\right)  _{V,\xi}.
\]

The relation between $K_{T}$ and $K_{T,\xi}$ and between $K_{S}$ and
$K_{S,\xi}$ are obtained by treating $V$ as a function of $T_{0},P_{0}$ and
$\xi$ and of $S,P_{0}$ and $\xi$, respectively. Using%
\begin{align}
dV  &  =\left(  \frac{\partial V}{\partial T_{0}}\right)  _{P_{0},\xi}%
dT_{0}+\left(  \frac{\partial V}{\partial P_{0}}\right)  _{T_{0},\xi}%
dP_{0}+\left(  \frac{\partial V}{\partial\xi}\right)  _{T_{0},P_{0}}%
d\xi,\label{V_T0_P0_Relation}\\
dV  &  =\left(  \frac{\partial V}{\partial S}\right)  _{P_{0},\xi}dS+\left(
\frac{\partial V}{\partial P_{0}}\right)  _{S,\xi}dP_{0}+\left(
\frac{\partial V}{\partial\xi}\right)  _{S,P_{0}}d\xi, \label{V_S_P0_Relation}%
\end{align}
we find that
\begin{equation}
K_{T}=K_{T,\xi}-\frac{1}{V}\left(  \frac{\partial V}{\partial\xi}\right)
_{T_{0},P_{0}}\left(  \frac{\partial\xi}{\partial P_{0}}\right)  _{T_{0}%
},\ \ K_{S}=K_{S,\xi}-\frac{1}{V}\left(  \frac{\partial V}{\partial\xi
}\right)  _{S,P_{0}}\left(  \frac{\partial\xi}{\partial P_{0}}\right)  _{S},
\label{K_K_Xi_Relation}%
\end{equation}
which is similar to similar relations for the heat capacity in Eq.
(\ref{C_C_Xi_Relation}). We similarly find that
\begin{equation}
\alpha_{P}=\alpha_{P,\xi}-\frac{1}{V}\left(  \frac{\partial V}{\partial\xi
}\right)  _{T_{0},P_{0}}\left(  \frac{\partial\xi}{\partial T_{0}}\right)
_{P_{0}}. \label{Alpha_Alpha_Xi_Relation}%
\end{equation}
Let us consider $\left(  \partial V/\partial P_{0}\right)  _{T_{0},\xi}$:%
\[
\left(  \frac{\partial V}{\partial P_{0}}\right)  _{T_{0},\xi}=\frac
{\partial(V,T_{0},\xi)}{\partial(P_{0},T_{0},\xi)}=\frac{\partial(V,S,\xi
)}{\partial(P_{0},S,\xi)}\frac{\partial(P_{0},S,\xi)}{\partial(P_{0},T_{0}%
,\xi)}\frac{\partial(V,T_{0},\xi)}{\partial(V,S,\xi)}=\left(  \frac{\partial
V}{\partial P_{0}}\right)  _{S,\xi}\frac{C_{P,\xi}}{C_{V,\xi}}.
\]
Similarly, we find that, see Sect. \ref{Marker_Subspace},
\[
\left(  \frac{\partial V}{\partial P_{0}}\right)  _{T_{0}}=\left(
\frac{\partial V}{\partial P_{0}}\right)  _{S}\frac{C_{P}}{C_{V}}%
\]
Thus, we have the standard identity for both kinds of compressibility:%
\begin{equation}
\frac{C_{P,\xi}}{C_{V,\xi}}\equiv\frac{K_{T,\xi}}{K_{S,\xi}},\ \ \frac{C_{P}%
}{C_{V}}\equiv\frac{K_{T}}{K_{S}} \label{Response_Ratios}%
\end{equation}

Let us again consider $K_{S,\xi}$. Rewriting%
\[
\left(  \frac{\partial V}{\partial P_{0}}\right)  _{S,\xi}=\frac
{\partial(V,S,\xi)/\partial(P_{0},T_{0},\xi)}{\partial(P_{0},S,\xi
)/\partial(P_{0},T_{0},\xi)}=\left(  \frac{\partial V}{\partial P_{0}}\right)
_{T_{0},\xi}-\frac{(\partial S/\partial P_{0})_{T_{0},\xi}(\partial V/\partial
T_{0})_{P_{0},\xi}}{(\partial S/\partial T_{0})_{P_{0},\xi}},
\]
we find that
\[
K_{T,\xi}\equiv K_{S,\xi}-\frac{(\partial S/\partial P_{0})_{T_{0},\xi
}(\partial V/\partial T_{0})_{P_{0},\xi}}{V(\partial S/\partial T_{0}%
)_{P_{0},\xi}}=K_{S,\xi}-(\partial S/\partial P_{0})_{T_{0},\xi}(\partial
V/\partial S)_{P_{0},\xi}..
\]

\section{\textbf{Prigogine-Defay Ratio}\label{Marker_PD_Ratio}}

Let us consider Figs. \ref{Fig_GlassTransition_V} and
\ref{Fig_GlassTransition_G} again that describe various kinds of glass
transitions: the apparent transitions at $T_{0\text{G}}$\ (point D) and
$T_{0\text{g}}^{(\text{A})}$\ (point C) and the conventional transitions at
$T_{0\text{G}}$\ (point D) and $T_{0\text{g}}$\ (point B). From the discussion
in Sect. \ref{Marker_Glass_Transitions}, we know that the Gibbs free energies
have a discontinuity between the two states involved at the apparent
transitions. Even the volumes and the entropies exhibit discontinuities at
these transitions. On the other hand, the Gibbs free energies, volumes and
entropies have no discontinuities at the conventional transitions at
$T_{0\text{G}}$ and $T_{0\text{g}}$ due to the continuity of the state. Let us
introduce the difference
\begin{equation}
\Delta q\equiv q_{\text{I}}-q_{\text{II}} \label{G_L_Difference}%
\end{equation}
for any quantity $q$ at a given $T_{0},P_{0}$ in the two possible states I and
II$.$ For the apparent glass transition at $T_{0\text{G}}$, $q_{\text{I}%
},q_{\text{II}}$ are the values of $q$ in GL and L, respectively, at
$T_{0\text{G}}$; for the apparent glass transition at $T_{0\text{g}%
}^{(\text{A})}$, $q_{\text{I}},q_{\text{II}}$ are the values of $q$ in gL and
L, respectively at $T_{0\text{g}}^{(\text{A})}$. For the conventional glass
transition at $T_{0\text{G}}$, $q_{\text{I}},q_{\text{II}}$ are the values of
$q$ in the glass GL and gL, respectively, at $T_{0\text{G}}$; for the
precursory glass transition at $T_{0\text{g}}$, $q_{\text{I}},q_{\text{II}}$
are the values of $q$ in gL and L, respectively at $T_{0\text{g}}$. These
states are summarized in the Table below.%
\[
\ \ \ \ \ \ \
\begin{tabular}
[c]{lllll}
&  & Table: Various & States & \\
& Apparent $T_{0\text{G}}$ & Apparent $T_{0\text{g}}^{(\text{A})}$ &
Conventional $T_{0\text{g}}$ & Conventional $T_{0\text{G}}$\\
I & \ \ \ \ \ \ GL & \ \ \ \ \ \ \ \ gL & \ \ \ \ \ \ \ \ \ gL &
\ \ \ \ \ \ \ \ \ \ GL\\
II & \ \ \ \ \ \ \ \ L & \ \ \ \ \ \ \ \ \ L & \ \ \ \ \ \ \ \ \ \ L &
\ \ \ \ \ \ \ \ \ \ \ gL
\end{tabular}
\]
\ \ \ \ \ \ \ \ \ \ \ \ \ \ \ \ \ \ \ \ \ \ \ \ \ \ \ \ \ \ \ \ \ \ \ \ \ \ \ \ \ \ \ \ \ \ \ 

In terms of the discontinuities $\Delta C_{P},\Delta K_{T}$ and $\Delta
\alpha_{P}$, the Prigogine-Defay ratio \cite{Prigogine-Defay} is traditionally
defined as
\cite{Prigogine-Defay,Davies,Goldstein,DiMarzio,Gupta,Gutzow-Pi,Nemilov,Garden}%

\[
\Pi^{\text{trad}}\equiv\frac{\Delta C_{P}\Delta K_{T}}{VT_{0}(\Delta\alpha
_{P})^{2}},
\]
where it is assumed that the volume is the same in both states at $T_{0}%
,P_{0}$, as is evident from earlier work. As we will see below, the volume is
normally not continuous at the apparent glass transitions, used in most
experimental and theoretical analyses of the glass transition. To allow for
this possibility, we will consider the following equivalent definition of the
Prigogine-Defay ration in this work:%
\begin{equation}
\Pi\equiv\frac{\Delta C_{P}\Delta K_{T}}{T_{0}(\Delta V\alpha_{P}%
)(\Delta\alpha_{P})}, \label{P-D_Ratio}%
\end{equation}
where we have absorbed $V$ in one of the $\Delta\alpha_{P}$-factors. It is
clear that $\Pi$ is not different from $\Pi^{\text{trad}}$ when the volume is
the same as happens for conventional transitions.

As the experimentalists have no control over the internal variables, and can
only manipulate the observables $\mathbf{X}$ by controlling the fields
$\mathbf{y}_{0}$ of the medium, we will discuss the evaluation of the
Prigogine-Defay ratio in the subspace of $\mathbf{y}_{0}$ of the complete
thermodynamic space of $\mathbf{y}_{0},\mathbf{a}_{0}$. We will consider the
simplest possible case in which the subspace reduces to the $T_{0}$-$P_{0}$
plane. Therefore, we will restrict ourselves to this plane in the following,
knowing very well that the GL and gL are also determined by the set
$\boldsymbol{\xi}$ of internal variables; see Sect. \ref{Marker_Subspace}. We
will consider the general case of several internal variables $\xi
_{k},k=1,2,\cdots,n$.

\subsection{Conventional Transitions at $T_{0\text{g}}$ and $T_{0\text{G}}$}

We will first consider the Prigogine-Defay ratio $\Pi_{\text{g}}$ at the
conventional transitions at points B and D (see Figs.
\ref{Fig_GlassTransition_V} and \ref{Fig_GlassTransition_G}). The continuity
of the state across B and D means that $E,V$ and $S$ remain continuous across
the conventional transitions at B and D. This is consistent with the
continuity of the Gibbs free energy. Let us first consider the transition at
B, where the relaxation time $\tau$ of the system becomes equal to the
observation time-scale $\tau_{\text{obs}}$, so that both states gL and L
remain in equilibrium with the medium. Thus, $T=T_{0},P=P_{0}$, and
$\mathbf{A}=\mathbf{A}_{0}=0$ for both states at B. Therefore, there is no
need to consider the internal variables in the Gibbs free energy, as they are
not independent variables. Moreover, $V=(\partial G/\partial P_{0})_{T_{0}}$
and $S=-(\partial G/\partial T_{0})_{P_{0}}$. Thus, the Gibbs free energy and
its derivatives with respect to $T_{0},P_{0}$ are continuous at B; the second
derivatives need not be. It is clear that B represents a point that resembles
a continuous transition in equilibrium; it turns into a glass transition curve
$T_{0\text{g}}(P_{0})$ of continuous transitions in the $T_{0}$ -$P_{0}$ plane.

For the transition at D, we have a glass GL on the low-temperature side, and
gL at the high temperature side; both states are out of equilibrium and have
the same temperature $T(t)$ and pressure $P(t)$, different from $T_{0},P_{0}$,
respectively at the transition. Similarly, $A(t)\neq0$ is the same in both
states. The important characteristics of the conventional transitions are the
continuity of $E,V$ and $S$ at B and D. We now follow the consequences of
these continuities.

\subsubsection{Continuity of Volume}

From the continuity of the volume, we have
\begin{equation}
d\Delta_{\text{g}}\ln V=\Delta_{\text{g}}\left(  \frac{\partial\ln V}{\partial
T_{0}}\right)  _{P_{0}}dT_{0}+\Delta_{\text{g}}\left(  \frac{\partial\ln
V}{\partial P_{0}}\right)  _{T_{0}}dP_{0}=0, \label{Volume_Continuity}%
\end{equation}
where $\Delta_{\text{g}}q$ denotes the difference in Eq. (\ref{G_L_Difference}%
) at the conventional glass transitions, and the derivatives are also
evaluated at the transition points. This equation can be written in terms of
the compressibilities and the expansion coefficients in the two states at the
glass transition temperature $T_{0\text{g}}$ or $T_{0\text{G}}$:%
\begin{equation}
\left.  \frac{dT_{0}}{dP_{0}}\right\vert _{\text{tr}}=\frac{\Delta_{\text{g}%
}K_{T}}{\Delta_{\text{g}}\alpha_{P}}; \label{Transition_slope_V_1}%
\end{equation}
the isothermal compressibility $K_{T}$ and the isobaric expansion coefficient
$\alpha_{P}$\ are given in Eqs. (\ref{Heat_Capacity}) and
(\ref{Expansion_Coefficients}), respectively, and can be expressed in terms of
the derivatives of the internal variable $\boldsymbol{\xi}$, such as given in
Eqs. (\ref{K_K_Xi_Relation}) and (\ref{Alpha_Alpha_Xi_Relation}) for a single
internal variable $\xi$. We make no assumption about these $\xi$-derivatives,
such as their vanishing or any assumption about freezing of $\xi$ at its value
at B; indeed, we expect $\xi$ to change continuously over BC. Of course, we
must remember that $\xi$ is an independent thermodynamic variable in gL and GL
states only, and not in the L state \cite{Gujrati-Non-Equilibrium-II}. The
slope equation (\ref{Transition_slope_V_1}) determines the variation of
$T_{0\text{g}}$ or $T_{0\text{G}}$ with the medium pressure $P_{0}$ along the
glass transition curve $T_{0\text{g,G}}(P_{0})$ in the $T_{0}$ -$P_{0}$ plane,
regardless of $\boldsymbol{\xi}.$ The form of the above equation does not
depend on the number of internal variables, provided we use the proper
definitions of $K_{T}$ and $\alpha_{P}$ as given in Eqs. (\ref{Heat_Capacity})
and (\ref{Expansion_Coefficients}), respectively. Its form follows from the
continuity of the volume at the conventional glass transition.

\subsubsection{Continuity of Entropy}

From the continuity of the entropy at $T_{0\text{g}}$, we similarly have
\begin{equation}
d\Delta_{\text{g}}S=\Delta_{\text{g}}\left(  \frac{\partial S}{\partial T_{0}%
}\right)  _{P_{0}}dT_{0}+\Delta_{\text{g}}\left(  \frac{\partial S}{\partial
P_{0}}\right)  _{T_{0\text{g}}}dP_{0}=0, \label{Entropy_Continuity}%
\end{equation}
from which we obtain at the precursory glass transition at B%
\begin{equation}
\left.  \frac{dT_{0\text{g}}}{dP_{0}}\right\vert _{\text{tr}}=\frac
{T_{0}\Delta_{\text{g}}(V\alpha_{P})}{\Delta_{\text{g}}C_{P}}=\frac
{V_{\text{g}}T_{0}\Delta_{\text{g}}\alpha_{P}}{\Delta_{\text{g}}C_{P}},
\label{Transition_slope_S_1}%
\end{equation}
where we have used the equilibrium Maxwell relation $(\partial S/\partial
P_{0})_{T_{0}}=-(\partial V/\partial T_{0})_{P_{0}}=V\alpha_{P}$; see Eq.
(\ref{Maxwell_Relations}) or Eq. (\ref{S_P0_V_T0_relation}) applied to this
case. Here $V_{\text{g}}$ is the common volume of gL and L at B and has been
taken out of $\Delta_{\text{g}}(V\alpha_{P})$. Again, this relation for the
slope is quite general, independent of the number of internal variables in gL
state at lower temperatures $T_{0}<T_{0\text{g}}$. Accordingly,%
\begin{equation}
\Pi_{\text{g}}\equiv\frac{\Delta_{\text{g}}C_{P}\Delta_{\text{g}}K_{T}%
}{V_{\text{g}}T_{0}(\Delta_{\text{g}}\alpha_{P})^{2}}=1,
\label{P-D_Ratio_Glass}%
\end{equation}
as expected for equilibrium states.\ It is a consequence of the glass
transition being a continuous transition between equilibrium states at B. As
we will see below, it is not merely a consequence of the continuity of volume
and entropy simultaneously.

Let us now consider the glass transition at $T_{0\text{G}}$. It follows from
Eq. (\ref{Heat_Capacity}) that
\[
\Delta_{\text{g}}\left(  \frac{\partial S}{\partial T_{0}}\right)  _{P_{0}%
}=\frac{\Delta_{\text{g}}C_{P}}{T}.
\]
In conjunction with Eq. (\ref{S_P0_V_T0_relation}), we find that
\[
\left.  \frac{dT_{0\text{G}}}{dP_{0}}\right\vert _{\text{tr}}=\frac
{V_{\text{G}}T\Delta_{\text{g}}\alpha_{P}}{\Delta_{\text{g}}C_{P}}%
\frac{\left(  \partial P/\partial P_{0}\right)  _{V}}{\left(  \partial
T/\partial T_{0}\right)  _{S}},
\]
where $V_{\text{G}}$ is the common volume of gL and GL at D and has been taken
out of $\Delta_{\text{g}}(V\alpha_{P})$. We finally obtain%
\begin{equation}
\Pi_{\text{G}}\equiv\frac{\Delta_{\text{g}}C_{P}\Delta_{\text{g}}K_{T}%
}{V_{\text{G}}T_{0}(\Delta_{\text{g}}\alpha_{P})^{2}}=\frac{T}{T_{0}}%
\frac{\left(  \partial P/\partial P_{0}\right)  _{V}}{\left(  \partial
T/\partial T_{0}\right)  _{S}}\neq1 \label{P-D_Ratio_Glass_Genuine}%
\end{equation}
for the conventional glass transition at D. The deviation of $\Pi_{\text{G}}$
from unity is independent of the number of internal variables. It will be
different from unity even if we have no internal variables.

\subsection{Apparent Glass Transitions at $T_{0\text{g}}^{(\text{A})}$ and
$T_{0\text{G}}$}

Unfortunately, it is not a common practice to determine the Prigogine-Defay
ratio at the conventional transitions at temperatures $T_{0\text{g}}(P_{0})$
or $T_{0\text{G}}(P_{0})$, which resemble continuous transition in that the
volume and entropy are continuous, along with the Gibbs free energy. In
experiments, one determines the ratio at apparent glass transitions either at
D or at $T_{0\text{g}}^{(\text{A})}(P_{0})$ in the glass transition region BD;
see Figs. \ref{Fig_GlassTransition_V} and \ref{Fig_GlassTransition_G}. In
these transitions, there are discontinuities in the $G,E,V$ and $S$. The
extrapolated point C (see Fig. \ref{Fig_GlassTransition_V}) identifies the
apparent glass transition temperature $T_{0\text{g}}^{(\text{A})}(P_{0})$,
which cannot be treated as a transition temperature because the Gibbs free
energy in the two states (gL and L) are not equal, as is clearly seen in Fig.
\ref{Fig_GlassTransition_G}. It is clear from Fig. \ref{Fig_GlassTransition_V}
that the volume is also different in gL and L at the apparent glass transition
$T_{0\text{g}}^{(\text{A})}(P_{0})$. The discontinuity of the volume should
not be confused with the continuity of the \emph{extrapolated} volumes used to
determine the location of the phenomenological glass transition $T_{0\text{g}%
}^{(\text{A})}(P_{0})$. The extrapolated glass volume does not represent the
physical volume of the glass at $T_{0\text{g}}^{(\text{A})}(P_{0})$ given by
the point on the curve BD in Fig. \ref{Fig_GlassTransition_V}. The
discontinuity is between the physical volumes of gL and L at $T_{0\text{g}%
}^{(\text{A})}(P_{0})$. We already know that both the entropy and the enthalpy
of the glass continue to decrease during vitrification as the system relaxes
\cite{Gujrati-Non-Equilibrium-I}. Indeed, the volume of the glass or gL also
relaxes towards that of the supercooled liquid L. This will also be true at
$T_{0\text{g}}^{(\text{A})}(P_{0})$ so that the volume and the entropy of gL
are higher than their values in the supercooled liquid at $T_{0\text{g}%
}^{(\text{A})}(P_{0})$ in a vitrification process. The same sort of
discontinuities also occur at D. In the following, we will take into account
these discontinuities in the volume and entropy in determining the
Prigogine-Defay ratio at the apparent glass transitions at $T_{0\text{G}%
}(P_{0})$ and $T_{0\text{g}}^{(\text{A})}(P_{0})$. The discontinuity of volume
$\Delta_{\text{g}}^{(\text{A})}V$ ($\neq0$) causes a modification of Eq.
(\ref{Volume_Continuity}) at these transitions:
\begin{equation}
\left.  \frac{dT_{0}}{dP_{0}}\right\vert _{\text{tr}}=\frac{\delta\ln
V_{P}^{(\text{A})}+\Delta_{\text{g}}^{(\text{A})}K_{T}}{\Delta_{\text{g}%
}^{(\text{A})}\alpha_{P}}=\frac{\Delta_{\text{g}}^{(\text{A})}K_{T}}%
{\Delta_{\text{g}}^{(\text{A})}\alpha_{P}}(1+\delta_{\text{g}}^{(\text{A}%
)}V_{P}) \label{Transition_slope_V_A}%
\end{equation}
in terms of
\begin{equation}
\delta\ln V_{P}^{(\text{A})}\equiv\left.  d\Delta_{\text{g}}^{(\text{A})}\ln
V/dP_{0}\right\vert _{\text{tr}} \label{lnV_Discontinuity}%
\end{equation}
at $T_{0\text{g}}^{(\text{A})}$ or $T_{0\text{G}}$, as the case may be; the
three $\Delta_{\text{g}}^{(\text{A})}$'s are the difference $\Delta$\ in Eq.
(\ref{G_L_Difference}) evaluated at $T_{0\text{g}}^{(\text{A})}$ or
$T_{0\text{G}}$, and the new quantity $\delta_{\text{g}}^{(\text{A})}V_{P}$
has an obvious definition:%
\begin{equation}
\delta_{\text{g}}^{(\text{A})}V_{P}=\frac{\delta\ln V_{P}^{(\text{A})}}%
{\Delta_{\text{g}}^{(\text{A})}K_{T}} \label{Volume_Correction_term}%
\end{equation}
at the appropriate temperature. This contribution would vanish under the
approximation $\Delta_{\text{g}}^{(\text{A})}\ln V\simeq0$, or $\delta\ln
V_{P}^{(\text{A})}\simeq0$. The slope equation (\ref{Transition_slope_V_A})
must always be satisfied at the apparent glass transition temperature. The
quantity $\delta\ln V_{P}^{(\text{A})}$\ in it represents the variation of the
discontinuity
\[
\Delta_{\text{g}}^{(\text{A})}\ln V=\ln V_{\text{I}}(T_{0},P_{0})-\ln
V_{\text{II}}(T_{0},P_{0})
\]
with pressure along the apparent glass transition curve $T_{0\text{g}%
}^{(\text{A})}(P_{0})$ or $T_{0\text{G}}(P_{0}),$ and can also be found
experimentally. Indeed, we can treat $\Delta_{\text{g}}^{(\text{A})}\ln V$ as
a function of $P_{0\text{g}}\equiv P_{0}(T_{0\text{g}}^{(\text{A})})$ along
the transition curves$.$ Then the contribution from the volume discontinuity
is given by
\begin{equation}
\delta\ln V_{P}^{(\text{A})}=\frac{1}{V_{\text{I}}}\left.  \frac{dV_{\text{I}%
}(P_{0})}{dP_{0}}\right\vert _{\text{tr}}-\frac{1}{V_{\text{II}}}\left.
\frac{dV_{\text{II}}(P_{0})}{dP_{0}}\right\vert _{\text{tr}}.
\label{lnV_derivative_difference}%
\end{equation}
We can use Eqs. (\ref{K_K_Xi_Relation}) and (\ref{Alpha_Alpha_Xi_Relation}) to
express the slope in terms of $\Delta_{\text{g}}K_{T,\xi}$ and\ $\Delta
_{\text{g}}\alpha_{P,\xi}$:%
\begin{equation}
\left.  \frac{dT_{0}}{dP_{0}}\right\vert _{\text{tr}}=\frac{\delta\ln
V_{P}^{(\text{A})}+\Delta_{\text{g}}^{(\text{A})}K_{T,\xi}-V_{\xi,\text{I}%
}\left.  \partial\xi/\partial P_{0}\right\vert _{\text{tr}}/V_{\text{I}}%
}{\Delta_{\text{g}}^{(\text{A})}\alpha_{P,\xi}-V_{\xi,\text{I}}\left.
\partial\xi/\partial T_{0}\right\vert _{\text{tr}}/V_{\text{I}}},
\label{Transition_slope_V_0}%
\end{equation}
where $V_{\xi,\text{G}}$ represents the derivative $\left(  \partial
V_{\text{I}}/\partial\xi\right)  _{T_{0},P_{0}}$, and $V_{\text{I}}$ is the GL
volume at $T_{0\text{G}}$ or the gL volume at $T_{0\text{g}}^{(\text{A})}$.
The $\xi$-contribution from the L state is absent due to the vanishing of the
affinity $\mathbf{A}_{0}(=0)$ in the L.

Let us now consider the differential of the entropy difference at the apparent
glass transition in the $T_{0}$ -$P_{0}$ plane:%
\[
d\Delta_{\text{g}}^{(\text{A})}S=\Delta_{\text{g}}^{(\text{A})}\left(
\frac{\partial S}{\partial T_{0}}\right)  _{P_{0}}dT_{0}+\Delta_{\text{g}%
}^{(\text{A})}\left(  \frac{\partial S}{\partial P_{0}}\right)  _{T_{0}}%
dP_{0},
\]
from which we find that
\begin{equation}
\left.  \frac{dT_{0}}{dP_{0}}\right\vert _{\text{tr}}=\frac{\delta
S_{P}^{(\text{A})}-\Delta_{\text{g}}^{(\text{A})}\left(  \partial S/\partial
P_{0}\right)  _{T_{0}}}{\Delta_{\text{g}}^{(\text{A})}\left(  \partial
S/\partial T_{0}\right)  _{P_{0}}}, \label{Transition_slope_S_A}%
\end{equation}
with%
\begin{equation}
\delta S_{P}^{(\text{A})}\equiv\left.  d\Delta_{\text{g}}^{(\text{A})}%
S/dP_{0}\right\vert _{\text{tr}}; \label{S_Discontinuity}%
\end{equation}
it represents the rate of variation of the entropy discontinuity
\[
\Delta_{\text{g}}^{(\text{A})}S=S_{\text{I}}(T_{0},P_{0})-S_{\text{II}}%
(T_{0},P_{0})
\]
along the apparent glass transition curves. Following the steps in deriving
Eq. (\ref{lnV_derivative_difference}), we find that the contribution from the
entropy discontinuity is given by
\begin{equation}
\delta S_{P}^{(\text{A})}=\left.  \frac{dS_{\text{I}}(P_{0})}{dP_{0}%
}\right\vert _{\text{tr}}-\left.  \frac{dS_{\text{II}}(P_{0})}{dP_{0}%
}\right\vert _{\text{tr}}. \label{S_derivative_difference}%
\end{equation}
\qquad

The derivative $\left(  \partial S_{\text{I}}/\partial P_{0}\right)  _{T_{0}}$
in the second term in the numerator in Eq. (\ref{Transition_slope_S_A}) can be
manipulated as in Eq. (\ref{S_P0_V_T0_relation}):
\[
\frac{\partial(S,T_{0})}{\partial(P_{0},T_{0})}=-\frac{\partial(S,T_{0}%
)}{\partial(V,P_{0})}\frac{\partial(P_{0},V)}{\partial(P_{0},T_{0})}=-\left(
\frac{\partial V}{\partial T_{0}}\right)  _{P_{0}}\frac{\partial(S,T_{0}%
)}{\partial(V,P_{0})},
\]
in which the last Jacobian reduces to unity under equilibrium by the use of
the Maxwell relation $\partial(V,P=P_{0})=$ $\partial(S,T=T_{0})$. We,
therefore, write%
\[
\frac{\partial(S,T_{0})}{\partial(V,P_{0})}=\frac{(\partial P/\partial
P_{0})_{V}}{(\partial T/\partial T_{0})_{S}}=1+\delta S_{VS}^{\text{I}}%
\]
for the glassy state; this equation also defines the modification $\delta
S_{VP}^{\text{I}}$ given by%
\begin{equation}
\delta S_{VS}^{\text{I}}=\frac{(\partial P/\partial P_{0})_{V}}{(\partial
T/\partial T_{0})_{S}}-1 \label{Entropy_Correction_term}%
\end{equation}
for the glassy state, where $T,P$ are the internal temperature, pressure of
the glassy state. It vanishes under the approximation $T=T_{0}$ and $P=P_{0}$.
We now have
\[
\left(  \frac{\partial S_{\text{I}}}{\partial P_{0}}\right)  _{T_{0}}=-\left(
\frac{\partial V_{\text{I}}}{\partial T_{0}}\right)  _{P_{0}}(1+\delta
S_{VS}^{\text{I}})=-V_{\text{I}}(1+\delta S_{VS}^{\text{I}})\alpha
_{P}^{\text{I}}.
\]
For the supercooled liquid, which represents an equilibrium state, we
evidently have ($S_{\text{II}}\equiv S_{\text{L}}$)%
\[
\left(  \frac{\partial S_{\text{L}}}{\partial P_{0}}\right)  _{T_{0}}=-\left(
\frac{\partial V_{\text{L}}}{\partial T_{0}}\right)  _{P_{0}}=-V_{\text{L}%
}\alpha_{P}^{\text{L}},
\]
so that
\[
\Delta_{\text{g}}^{(\text{A})}\left(  \partial S/\partial P_{0}\right)
_{T_{0}}=-\Delta_{\text{g}}^{(\text{A})}(V\alpha_{P})-V_{\text{I}}\alpha
_{P}^{\text{I}}\delta S_{VS}^{\text{I}}.
\]

We now turn to the denominator in Eq. (\ref{Transition_slope_S_A}). For the
supercooled liquid state, whose temperature is $T_{0}$, we have%
\[
\left(  \frac{\partial S_{\text{L}}}{\partial T_{0}}\right)  _{P_{0}}%
=\frac{C_{P}^{\text{L}}}{T_{0}};
\]
we must use $T_{0\text{g}}^{(\text{A})}$ or $T_{0\text{G}}$\ for $T_{0}$ to
evaluate this slope at the appropriate apparent glass transition. For the
glass, whose internal temperature is $T$, we have%
\[
\left(  \frac{\partial S_{\text{I}}}{\partial T_{0}}\right)  _{P_{0}}%
=\frac{C_{P}^{\text{I}}}{T}\equiv(1+\delta T^{\text{I}})\frac{C_{P}^{\text{I}%
}}{T_{0}},
\]
where we have introduced a correction parameter%
\begin{equation}
\delta T^{\text{I}}\equiv T_{0\text{g,G}}^{(\text{A})}/T-1,
\label{Temperature_Correction_term}%
\end{equation}
with $T_{0\text{g,G}}^{(\text{A})}$ denoting $T_{0\text{g}}^{(\text{A})}$ or
$T_{0\text{G}}$\ as the case may be. Again, this modification term vanishes
under the approximation $T=T_{0}$. We thus find that
\[
T_{0}\Delta_{\text{g}}^{(\text{A})}\left(  \partial S/\partial T_{0}\right)
_{P_{0}}=\Delta_{\text{g}}^{(\text{A})}C_{P}+C_{P}^{\text{I}}\delta
T^{\text{I}}.
\]
Equating the two different versions of the slope in Eqs.
(\ref{Transition_slope_V_A}) and (\ref{Transition_slope_S_A}), we have%
\begin{align*}
\frac{\Delta_{\text{g}}^{(\text{A})}K_{T}}{\Delta_{\text{g}}^{(\text{A}%
)}\alpha_{P}}(1+\delta_{\text{g}}^{(\text{A})}V_{P})  &  =T_{0\text{g,G}%
}^{(\text{A})}\frac{\delta S_{P}^{(\text{A})}+\Delta_{\text{g}}^{(\text{A}%
)}(V\alpha_{P})+V_{\text{I}}\alpha_{P}^{\text{I}}\delta S_{VS}^{\text{I}}%
}{\Delta_{\text{g}}^{(\text{A})}C_{P}+C_{P}^{\text{I}}\delta T^{\text{G}}}\\
&  \equiv\frac{T_{0\text{g,G}}^{(\text{A})}\Delta_{\text{g}}^{(\text{A}%
)}(V\alpha_{P})}{\Delta_{\text{g}}^{(\text{A})}C_{P}}(1+\delta^{\prime}%
\Pi_{\text{gA}}),
\end{align*}
where we have introduced a new quantity $\delta^{\prime}\Pi_{\text{gA}}$,
whose definition is obvious from the equality.

We finally find that the Prigogine-Defay ratio is given by
\begin{equation}
\Pi_{\text{gA}}\equiv\frac{\Delta_{\text{g}}^{(\text{A})}C_{P}\Delta
_{\text{g}}^{(\text{A})}K_{T}}{T_{0\text{g,G}}^{(\text{A})}\Delta_{\text{g}%
}^{(\text{A})}\alpha_{P}\Delta_{\text{g}}^{(\text{A})}(V\alpha_{P})}%
\equiv1+\delta\Pi_{\text{gA}}\equiv\frac{1+\delta^{\prime}\Pi_{\text{gA}}%
}{1+\delta_{\text{g}}^{(\text{A})}V_{P}} \label{P-D_Ratio_Glass_Apparent}%
\end{equation}
at the apparent glass transition. Its complete form is given by%
\begin{equation}
\Pi_{\text{gA}}=\frac{1+\left(  V_{\text{I}}\alpha_{P}^{\text{I}}\delta
S_{VS}^{\text{I}}+\delta S_{P}^{(\text{A})}\right)  /\Delta_{\text{g}%
}^{(\text{A})}(V\alpha_{P})}{(1+C_{P}^{\text{I}}\delta T^{\text{I}}%
/\Delta_{\text{g}}^{(\text{A})}C_{P})(1+\delta_{\text{g}}^{(\text{A})}V_{P})}.
\label{P-D_Ratio_Glass_Apparent_0}%
\end{equation}
It should be obvious that the Prigogine-Defay ratio is itself a function of
time as it depends on time-dependent quantities such as $\Delta_{\text{g}%
}^{(\text{A})}S$, $\delta T^{\text{I}},$ etc.

\subsubsection{Approximation A}

Let us assume that the discontinuities in the volume and entropy are
negligible or that the contributions $\delta\ln V_{P}^{(\text{A})}$ and
$\delta S_{P}^{(\text{A})}$\ are negligible. In that case, the Prigogine-Defay
ratio reduces to%
\[
\Pi_{\text{gA}}\simeq\frac{1+V_{\text{I}}\alpha_{P}^{\text{I}}\delta
S_{VS}^{\text{1}}/\Delta_{\text{g}}^{(\text{A})}(V\alpha_{P})}{1+C_{P}%
^{\text{I}}\delta T^{\text{I}}/\Delta_{\text{g}}^{(\text{A})}C_{P}},
\]
and will have a value different than $1$. Thus, the continuity of volume and
entropy alone is not sufficient to yield $\Pi_{\text{gA}}=1$, as noted above.
If we further approximate $T\simeq T_{0}$ and $P\simeq P_{0}$, then $\delta
S_{VS}^{\text{I}}\simeq0$ and $\delta T^{\text{I}}\simeq0$, and we obtain
$\Pi_{\text{gA}}\simeq1$. This is expected as the approximations change the
apparent glass transition into a continuous transition. If, however, we only
assume $P\simeq P_{0},$ but allow $T$ to be different from $T_{0}$, then%
\[
\delta S_{VS}^{\text{I}}\simeq\frac{1}{(\partial T/\partial T_{0}%
)_{S_{\text{I}}}}-1,
\]
and we still have $\Pi_{\text{gA}}\neq1$.

\subsubsection{Approximation B}

We make no assumption about $\delta\ln V_{P}^{(\text{A})}$ and $\delta
S_{P}^{(\text{A})}$, but approximate $T\simeq T_{0}$ and $P\simeq P_{0}$. In
this case, $\delta S_{VS}^{\text{I}}\simeq0$ and $\delta T^{\text{I}}\simeq0$,
and we obtain%
\[
\Pi_{\text{gA}}\simeq\frac{1+\delta S_{P}^{(\text{A})}/\Delta_{\text{g}%
}^{(\text{A})}(V\alpha_{P})}{1+\delta_{\text{g}}^{(\text{A})}V_{P}}.
\]
If, however, the approximation $T\simeq T_{0}$ is not valid, we have%
\[
\Pi_{\text{gA}}\simeq\frac{1+\delta S_{P}^{(\text{A})}/\Delta_{\text{g}%
}^{(\text{A})}(V\alpha_{P})}{(1+C_{P}^{\text{I}}\delta T^{\text{G}}%
/\Delta_{\text{g}}^{(\text{A})}C_{P})(1+\delta_{\text{g}}^{(\text{A})}V_{P}%
)}.
\]
In both cases, $\Pi_{\text{gA}}\neq1$.

\subsection{Comparison with Other Attempts for $\Pi$}

As far as we know, almost all previous attempts
\cite{Prigogine-Defay,Davies,Goldstein,DiMarzio,Gupta,Gutzow-Pi,Nemilov,Garden}
in the evaluation of $\Pi$ are based on treating the glass transition as a
\emph{direct} transition from L to GL; the structure is supposed to be almost
frozen in the latter. As we see from Figs. \ref{Fig_GlassTransition_V} and
\ref{Fig_GlassTransition_G}, this can only occur at C between L and the
extrapolated branch DC. At C, there will be a discontinuity between the values
of the internal variable $\xi$; it will take the equilibrium value
$\xi_{\text{C}}^{\text{eq}}$ in L, but will take a non-equilibrium value
$\xi_{\text{C}}^{\text{etra}}\neq\xi_{\text{C}}^{\text{eq}}$ obtained along
DC. Similarly, $A=A_{0}=0$ in L at C, while $A=A_{\text{C}}\neq0$ in the
extrapolated GL at C. As C is obtained by matching the volumes, the volume
remains continuous, but there is no reason to believe that the entropy will
remain continuous in this transition. The Gibbs free energy obviously remains
discontinuous in this transition.

However, we have been careful in not treating this transition as an apparent
transition above for the simple reason that there is no guarantee that the
branch DC can be described by vitrification thermodynamics at the constant
cooling rate $r$. To see it most easily, we observe that as the cooling rate
is gradually taken to be slower and slower, the transition point B gradually
moves towards C along BF. However, the analog of BD will most certainly not be
identical to DC for the simple reason that the state of L will continuously
change to gL so that the values of $\xi$ and $A$ in gL at C will be identical
to their values $\xi=$ $\xi_{\text{C}}^{\text{eq}}$\ and $A=0$ in L at C.
Moreover, there is no guarantee that the extrapolated branch DC can even
satisfy thermodynamics with known controllable parameters $T_{0},P_{0}$ and
$r$. To treat this "transition" as a glass transition requires some
approximation, which we have avoided.

\section{Conclusions}

We have followed the consequences of internal equilibrium to derive
generalizations of equilibrium thermodynamic relations such as Maxwell's
relations, Clausius-Clapeyron relation, relations between response functions
(heat capacities, compressibilities, etc.) to non-equilibrium systems.
Non-equilibrium states are described not only by internal fields (temperature,
pressure, etc.) that are different from the medium, but also described by
internal variables which cannot be controlled from outside by the observer.
The observer can only control the observables. Thus, in this work, we have
also discussed how the thermodynamics should be described in the subspace of
the observables only. As glasses are a prime example of non-equilibrium
states, we have reviewed the notion of the glass transition. The frozen
structure known as the glass (GL) does not emerge directly out of the
equilibrium supercooled liquid (L). There is an intermediate non-equilibrium
state (gL) that is not yet frozen when it emerges continuously out of the
equilibrium liquid L. At a lower temperature, this state continuously turns
into GL. Because of this, we find that there is no one unique non-equilibrium
transition. We introduce four of the most conceptually useful transitions. At
two of them, which we term conventional glass transitions, the Gibbs free
energies and the states are continuous. Thus, they are the non-equilibrium
analog of the conventional continuous or second order transition between
equilibrium states. At the other two glass transitions, which we term apparent
glass transition, not only the states but also the Gibbs free energies are
discontinuous. Because of this, these transitions are examples of a zeroth
order transition where the free energy is discontinuous. But there is no
transition in the system itself at the apparent glass transition as discussed
in Sect. \ref{Marker_Glass_Transitions}.

We briefly review the use of Jacobians which are found extremely useful in
obtaining the generalization of the Maxwell relations. There are many other
Maxwell relations than reported here; they can be easily constructed. We then
discuss various response functions and obtain relationship between them in
non-equilibriums states. Surprisingly, many of these relations look similar in
form to those found in equilibrium thermodynamics.

We finally evaluate the Prigogine-Defay ratio at the four possible glass
transitions. We find that the ratio is normally different than $1$, except at
the conventional glass transition at the highest temperature, where it is
always equal to $1$, regardless of the number of internal variables. We also
find that the continuity of volume and entropy is not a guarantee for $\Pi=1$.
We compare our analysis of $\Pi$ with those carried out by other workers.

\begin{acknowledgement}
P.P. Aung was supported by a summer internship from NSF through the University
of Akron REU\ site for Polymer Science.
\end{acknowledgement}

\appendix{}

\section{Relation between 2- and 3-Jacobians}

Let us consider a function $F(x,y,z)$, where $x,y,z$ may stand for
$T_{0},P_{0},\xi$, respectively. Then%
\[
dF=F_{x,yz}dx+F_{y,zx}dy+F_{z,xy}dz,
\]
where we have used the compact notation%
\[
F_{x,yz}\equiv\left(  \frac{\partial F}{\partial x}\right)  _{yz},
\]
etc. Now,%
\[
\left(  \frac{\partial F}{\partial x}\right)  _{y}=F_{x,yz}+F_{z,xy}\left(
\frac{\partial z}{\partial x}\right)  _{y}.
\]
Similarly,
\[
\left(  \frac{\partial K}{\partial y}\right)  _{x}=K_{y,zx}+K_{z,xy}\left(
\frac{\partial z}{\partial y}\right)  _{x}.
\]
We express $F_{x,yz}$ as a $3$-Jacobian and manipulate it as follows:%
\begin{align*}
\frac{\partial(F,y,z)}{\partial(x,y,z)}  &  =\frac{\partial(F,y,z)}%
{\partial(K,x,z)}\frac{\partial(K,x,z)}{\partial(x,y,z)}=\frac{\partial
(F,y,z)}{\partial(K,x,z)}\left[  -\left(  \frac{\partial K}{\partial
y}\right)  _{x}+K_{z,xy}\left(  \frac{\partial z}{\partial y}\right)
_{x}\right] \\
&  =-\left(  \frac{\partial K}{\partial y}\right)  _{x}\left[  \frac
{\partial(F,y,z)}{\partial(K,x,z)}-K_{z,xy}\left(  \frac{\partial z}{\partial
K}\right)  _{x}\frac{\partial(F,y,z)}{\partial(K,x,z)}\right]  .
\end{align*}
Using this, we find that%
\[
\left(  \frac{\partial F}{\partial x}\right)  _{y}=-\left(  \frac{\partial
K}{\partial y}\right)  _{x}\left[  \frac{\partial(F,y,z)}{\partial
(K,x,z)}-K_{z,xy}\left(  \frac{\partial z}{\partial K}\right)  _{x}%
\frac{\partial(F,y,z)}{\partial(K,x,z)}+F_{z,xy}\frac{\partial(z,y)}%
{\partial(K,x)}\right]  .
\]
Let us call the quantity in the square brackets $D$, which can be rewritten as%
\[
D\equiv\frac{\partial(F,y)}{\partial(K,x)}D^{\prime},
\]
where
\[
D^{\prime}=\frac{\partial(F,y,z)}{\partial(K,x,z)}\frac{\partial
(K,x)}{\partial(F,y)}-K_{z,xy}\frac{\partial(z,x)}{\partial(F,y)}%
\frac{\partial(F,y,z)}{\partial(K,x,z)}+F_{z,xy}\frac{\partial(z,y)}%
{\partial(F,y)}.
\]
Using Eq. (\ref{Permutation_Property_1}), it can now be shown in a
straight-forward manner that%
\[
D^{\prime}=1,
\]
which proves that
\[
\left(  \frac{\partial F}{\partial x}\right)  _{y}=-\left(  \frac{\partial
K}{\partial y}\right)  _{x}\frac{\partial(F,y)}{\partial(K,x)},
\]
the desired result.

While we considered $F$ and $K$ as a function of $3$ variables, we can
generalize the result to any number of variables. We will not pause here to do that.

\end{document}